\documentclass{aa}
\usepackage[varg]{txfonts}
\usepackage[utf8]{inputenc}
\usepackage{graphicx}
\usepackage{array}
\usepackage{hyperref}
\usepackage{multirow}
\usepackage{natbib}
\usepackage{enumitem}
\usepackage{pifont}
\usepackage{xfrac}
\usepackage{nicefrac}
\usepackage{mathtools}
\usepackage{amsmath}

\usepackage{xcolor}

\usepackage{ulem}

\hypersetup{
    colorlinks=true,
    linkcolor=blue,
    citecolor=blue
}

\begin{document}

\title{On the coupled origin of the stellar IMF and multiplicity.}
\subtitle{The influence of hierarchical fragmentation on the Core Mass Function.}

\author{Thomasson B.$^{1}$
\and Joncour I.$^{1}$
\and Moraux E.$^{1}$ 
\and Motte F.$^{1}$
\and Yoo T.$^{2}$
\and Ginsburg A.$^{2}$}

\institute{Univ. Grenoble Alpes, CNRS, IPAG, 38000 Grenoble, France
\and Department of Astronomy, University of Florida, P.O. Box 112055, Gainesville, FL 32611-2055, USA}

\date{Received <date> /
Accepted <date>}

\abstract {
In the solar neighborhood, the Initial Mass Function (IMF) follows {is canonically} described by the Salpeter power-law slope for the high-mass range. As stars inherit their mass from their environment, their IMF may directly result from {a} Core Mass Function (CMF) through accretion, gravitational collapse, and fragmentation. This inheritance implies that the mass of the gaseous fragments may be connected to the properties of clustered and multiple stellar systems. In these systems, mass and multiplicity are related as more massive primaries are observed more frequently in multiple systems.}{
We aim to (i) quantify the influence of hierarchical fragmentation of cores on the resulting IMF, and (ii) determine the consequences of this fragmentation on the multiplicity of the stellar systems.}{
We employed a scale-free, hierarchical fragmentation model to investigate the stochastic fragmentation of the 2.5~kAU cores of the W43-MM2\&MM3 molecular cloud, whose CMF is top-heavy. We {also used this model} to quantify the influence of deterministic mass-dependent fragmentation processes.}{
Hierarchical fragmentation of gas clumps shifts the CMF towards lower mass range and can modify its shape. The shift is quantified by both the number of fragments produced at each level of fragmentation and the mass the fragments inherit from their parental core. Starting from the top-heavy power-law CMF observed in W43-MM2\&MM3, we show that at least four levels of hierarchical fragmentation are required to generate the turn-over peak of the cIMF. Within a radius of 0.2-2.5~kAU, massive stars ($M > 10~M_\odot$) have on average 0.9 companions, five times fewer than low-mass stars ($M < 0.1~M_\odot$); the latter are less dynamically stable and should disperse. We show that a universal IMF can emerge from mass-dependent fragmentation processes provided that more massive cores produce less fragments compared to lower mass cores and transfer their mass less efficiently to their fragments.}{{Hierarchical fragmentation alone cannot reconcile a universal IMF with observed stellar multiplicity. We propose that fragmentation is not scale-free but operates in two distinct regimes: a mass-dependent phase establishing the Salpeter slope and a mass-independent phase setting the turn-over.} {Our framework provides} a way to compare core subfragmentation in various star-forming regions and numerical simulations.
}


\maketitle
%
%

\section{Introduction}
\label{Section:Introduction}

The stellar Initial Mass Function (IMF) {is usually interpreted to be a} the probability density function (PDF) that describes the mass distribution of newly born stars. It is a diagnostic tool widely used by astronomers to assess star formation properties. The mathematical representation of the IMF has evolved over the years. \cite{Salpeter1955} first parameterised the intermediate and high-mass regime ($0.4 \le M \le 10~{\rm M}_\odot$) for field stars as a power-law, for which the fraction of stars $\zeta (M)$ per unit of logarithmic mass is given by $ \zeta(M) = {\rm d} N / {\rm d} \log M \propto M^{\Gamma} $ with $\Gamma = -1.35$, the so-called "Salpeter slope". However, approximating the IMF with a single power-law appeared to be too simple to represent the low-mass stars range ($M < 1~{\rm M}_\odot$). Other functions were proposed such as a log-normal function \citep{Miller1979, Scalo1986}, a combination of segmented broken-powerlaws \citep{kroupa2001} or even a combination of log-normal and power-law \citep{chabrier2003}. Non-segmented functions have also been introduced to describe the IMF shape \citep{demarchi2005,masch2013} with the benefit to give a smooth, continuous and more practical description of the IMF.

Despite their mathematical differences, all these functions share common shape properties to describe the seemingly universal distribution of stellar mass in the Milky Way (MW): (i) the Salpeter power-law above one solar-mass star ($M>1~{\rm M}_\odot$), (ii) a log-normal shape below $1~{\rm M}_\odot$ with a turn-over mass, also called peak, around $M$ {$\approx$} $ 0.1-0.4~{\rm M}_\odot$ defining a characteristic mass for stars, (iii) a low-mass cutoff in the brown dwarf regime around $M \approx 0.2~{\rm M}_\odot$ \citep{thies2015} and a high-mass cutoff around $M \approx 100-120~{\rm M}_\odot$ \citep{weidnerkroupa2004, Figer2005, Zinnecker2007, Kroupa2013, Tan2014}. Because of the statistical consistency, obtained over decades, of a unique shape of the IMF within resolved stellar populations in the disk of the MW, it has been proposed that the IMF may be universal and independent of time and space \cite[see e.g., the review of][]{Krumholz2014, lee2020, hennebelle2024}. Hereafter we refer such IMF {within the MW} as a canonical IMF (cIMF). {See \citealt{kroupa2024, gjergo2025} for further discussion regarding the IMF universality.}

Theoretical works were performed to interpret the characteristics describing the shape of the cIMF. First, the log-normal low-mass regime can result from random multiplicative fragmentation processes \citep{larson1973_fragmentation, elmegreen1983} but may also be inherited from the underlying gas density distribution of the parental cloud \citep{H&C2008}. Second, the turnover mass depends on the physical properties of the molecular cloud that govern the structure and the dynamics of the cloud via (i) the local thermal Jeans length \citep{Larson1998, Larson2005}, (ii) the supersonic turbulent pressure, (iii) the Mach number in molecular clouds \citep{padoan2002,padoan1997, hopkins2013_cmfvariation} and (iv) the equation of state of the gas \citep{Lee&H2018_adiafirstlarson}. Lastly, the scale-free nature of supersonic turbulence and gravity shaping the structure of molecular clouds predicts power-law behavior for the mass distribution of prestellar cores \citep{H&C2008, hopkins2013_cmfvariation, Guszejnov2016_turbulentfragIMF} while dynamical mass growth/depletion such as outflows \citep{Adams1996} or competitive accretion \citep{bonnell2001} can also result in a power-law IMF.

Nonetheless, possible variations of the IMF with the physical properties of the parental cloud, is still an open field of debates \cite[e.g.,][]{kroupa2001, Elmegreen2004, Weidner2006, Kroupa2013, Hopkins2018}. Some studies claim to have observed IMF variations, specifically within starburst clusters and globular clusters \cite[see also for a review,][]{Scalo1998, Scalo2005}. A number of effects such as the limited statistical sampling, sample incompleteness, dynamical evolution, stellar evolution uncertainties and unresolved multiple stellar systems may explain part of these variations. Such claims about IMF variations within the MW were critically reviewed to suggest that they do not appear statistically significant \citep{bastian2010_review}. More recently, there has been more robust evidence for `top-heavy' IMF, characterised by an excess of high-mass stars ($M>1~{\rm M}_\odot$ range) compared to the cIMF, in dwarfs galaxies \citep{gennaro2018a} and within the MW, for example, in the young nuclear cluster \citep{Lu2013} or in the Arches \citep{hosek2019}. {In addition, massive early-type galaxies appear to host bottom-heavy IMF \citep{conroy2017}, while their high metallicities require the IMF to be top-heavy \citep{yanjabkov2021}.} These findings raise questions about the physical conditions that allow a cloud to form stars with top-heavy or Salpeter-like IMFs. Therefore, the shape of the IMF may be the consequence of the underlying star formation processes that depend on the physical properties of the hosting cloud.


On the other hand, the mass function of prestellar cores, called the core mass function (CMF), observed in nearby star-forming regions resembles the cIMF as its high-mass tail can also be described by a power-law, close to the Salpeter one \citep{Motte1998, konyves2015, sokol2019, konyves2020, suarez2021}. Although recent works point out to the potential lack of robustness of CMF construction due to 2D spatial projection effects and spatial resolution limitation \citep{Louvet2021, Padoan2023}, a possible interpretation of the close resemblance between the CMF and the cIMF is that each core may in fact produce  a single star with a constant mass conversion efficiency \citep{alves2007}. In massive star-forming regions imaged by the ALMA-IMF large program (among which W43-MM1 and W43-MM2\&3), CMFs were recently observed with an excess of massive objects compared to the regular cIMF, resulting in shallower power-law indices at high-mass \citep{motte_unexpectedly_2018, pouteau2022, Nony2023, louvet2024}. Constant {star formation} efficiency {(i.e. the fraction of core mass that collapses into a star)} alone is not sufficient to recover the cIMF as it simply induces a translation of the CMF towards lower masses. Without a star formation mechanism to recover the cIMF, the existence of such top-heavy CMF challenges the seemingly universal IMF within our galaxy and supports its theoretical dependence on the parental gaseous environment. 

{To investigate the hypothesis that the stellar IMF emerges from a pre-existing CMF, two approaches can be considered. One may start from an observed IMF and reconstruct a synthetic CMF using clustering algorithms (e.g., \citealt{zhou2025}), or start with a population of prestellar cores and map their CMF with the cIMF through collapse or fragmentation (e.g., \citealt{hopkins2012_last_crossing}). In this work, we adopt the latter to investigate the hypothesis that the shape of the IMF emerges from the CMF.} To perform such mapping, one monitors the evolution of the mass distribution of gas structures until star formation {ends}, under the requirement that the cores should not sub-fragment but collapse into a single star with a given mass efficiency, without mergers. At this stage, the CMF of these non-fragmenting cores is supposed to be comparable to their stellar IMF in a one-to-one mapping. If the mass efficiency is not constant but depends on individual cores, or if the cores sub-fragment, this observational correspondence is not guaranteed. On the contrary, top-heavy CMF could yield the cIMF accounting for the suited fragmentation efficiencies, although the impact of these effects has never been quantified analytically.

Under the hypothesis that the cIMF emerges from the CMF, we expect fragmentation to determine both the spatial distribution and the multiplicity of the stellar systems formed within the cores. Multiplicity studies show that most stars possess a companion as they are structured in binaries, triples, or higher order systems \citep{lada2003efficiency, marksnkroupa2011, duchene_stellar_2013}. The proportion of multiples compared to isolated objects tends to increase with the mass of the most massive object contained in these systems. More than 80\% of the most massive {field} stars ($M > 10~\text{M}_\odot$) have at least one companion, while more than 80\% of the low-mass stars ($M < 0.1~\text{M}_\odot$) are single (see review of \citealt{offner2023} and \citealt{kroupa2024}). These observations are supported by numerical simulations \citep{bate2012, guzs2017_multiplicity} establishing that most massive protostars form with at least one gravitationally bound companion. Other simulations, in which core fragmentation weakly depends on core mass, succeeded to reconcile these multiplicity properties with the global shape of the cIMF \citep{houghton2024}.

{In this work, we investigate whether the hierarchical fragmentation of $< 0.01$~pc prestellar cores described by a CMF can shape both the cIMF and the observed stellar multiplicity. Although supersonic turbulence is inherently scale-free, \citet{thomasson2024} showed that below $0.1$~pc, turbulence becomes subsonic and fragmentation is instead driven by thermal pressure, resulting in a scale-free process down to the formation of the first Larson core \citep{larson1969_fcore}. We reintroduce the mathematical framework developed by \citet{thomasson2024} and apply it to the top-heavy CMF of \citet{pouteau2022}, composed of $\approx 0.01$ pc cores, following a scale-free fragmentation. We assign ad-hoc numerical values to the model parameters, in order to investigate the consequences of different fragmentation scenarios regardless of the underlying physical process driving fragmentation. For clarity, all acronyms and variable names used in this work are listed in Appendix~\ref{appendix:acronymTable}, in Tables~\ref{tab:acronyms} and~\ref{tab:definitions}.}

We introduce in Sect.~\ref{sec:Model of hierarchical fragmentation} our scale-free model of fragmentation that describes the successive hierarchical sub-fragmentation at different spatial scales within a cloud and quantify the variations of a CMF under hierarchical fragmentation as a function of spatial scale. In Sect.~\ref{sec:Hierarchical fragmentation applied to a top-heavy CMF}, we monitor the spatial evolution of the top-heavy CMF from W43-MM2\&MM3 region \citep{pouteau2022} described by a power-law distribution of index $\Gamma = -0.95$ in order to evaluate the influence of stochastic processes and mass repartition between fragments. We also discuss the conditions under which this top-heavy CMF fragments into the cIMF and its implications for the resulting stellar multiplicities. Next, in Sect.~\ref{sec:Impact of mass dependencies on a mass distribution}, we use this model to quantify the slope variations of any CMF considering different mass-dependent fragmentation prescriptions. Finally, in Sect.~\ref{sec:discussion} we discuss the properties that hierarchical fragmentation requires in order to reconcile both the general shape of the cIMF and the multiplicity observed in stellar systems, before concluding in Sect.~\ref{sec:conclusion}.

%
%
\section{Model of hierarchical fragmentation}
\label{sec:Model of hierarchical fragmentation}


\subsection{General framework}
\label{sec:General framework}

\begin{figure}
    \centering
    \includegraphics[width=7cm]{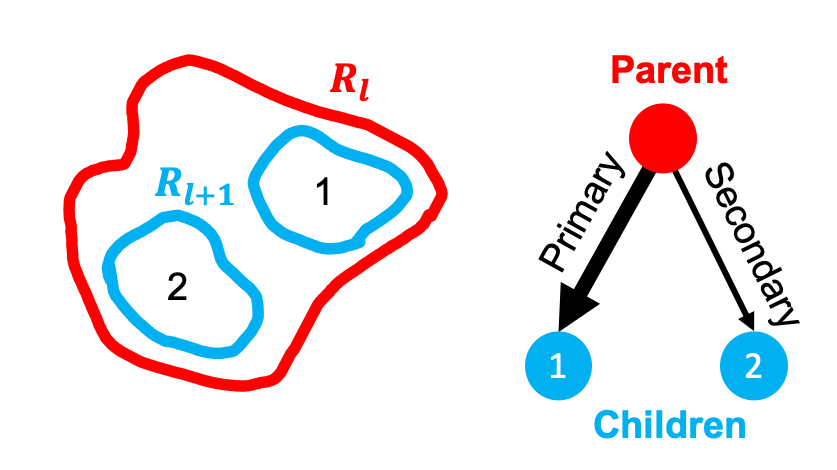}
    \caption{Sketch of the hierarchical fragmentation model used. A parental object at spatial scale $R_l$ (red) fragments into a number $n_l$ of children at scale $R_{l+1}$ (blue), here $n_l = 2$. The children are identified by an index $i$ where $i = 1$ represents the primary, (i.e. most massive) child.}
    \label{fig:sketchfrag}
\end{figure}

To investigate the influence of hierarchical fragmentation on the shape of a CMF, we aim to fragment the dense cores constituting this CMF and evaluate the resulting mass function. We use the hierarchical fragmentation model developed by \citet{thomasson2024}, which characterises the successive fragmentations of a dense gas structures across spatial scales. Fragments of size $R_{l+1}$ occupy a fragmentation level $l+1$ and are embedded within larger parental structures populating the level $l$ associated to scale $R_l > R_{l+1}$ (Fig.~\ref{fig:sketchfrag}). This hierarchical process is stochastic. {Stochasticity reflects the challenge of mapping a continuous gas distribution onto a discrete stellar population, which ultimately introduces an incomplete knowledge of the precise outcome of the number of fragments between two spatial scales.} The number of children $n_l$ produced by one parent at the level $l$ is random and follows a discrete probability distribution $p_l(n_l)$ which assigns, for each alternative $n_l$, a probability $p_l$ such that we can write its expected value $\bar{n_l}$ as

\begin{equation}
\bar{n_l} = \displaystyle \sum_{\{n_l\}} p_l n_l
\label{eq:expecancy_pl}
.\end{equation}

Let define as $\epsilon_l$ the mass efficiency corresponding to the ratio between the total mass of the $n_l$ children and the mass of their parent $M_l$ so

\begin{equation}
    \epsilon_l = \dfrac{\displaystyle \sum_{i=1}^{n_l} M_{l+1,i}}{M_{l}}
,\end{equation}

\noindent where $M_{l+1,i}$ is the mass of the $i-th$ child produced by one parent, with $i \leq n_l$ (Fig. \ref{fig:sketchfrag}). Once one parent produced its $n_l$ children, the latter need to share their parental mass reservoir $\epsilon_l M_l$. Each child may inherit different mass fraction from this common reservoir. The $i-th$ child produced inherits a mass

\begin{equation}
     M_{l+1,i} = M_{l} \epsilon_l \psi_{l,i}
     \label{eq:mass_of_children}
,\end{equation}

\noindent where $\psi_{l,i}$ is the fraction of mass the $i-th$ child inherits from the parental mass reservoir. In the following we consider a mass partition function that results in the formation of one dominant, more massive fragment and other less massive (Fig.~\ref{fig:sketchfrag}), or otherwise equally massive, satellite fragments

\begin{equation}
    \psi_{l,i} = 
    \begin{cases}
        \dfrac{q}{q + n_l - 1} & \text{if } i = 1 \\[15pt]
        \dfrac{1}{q + n_l - 1} & \text{otherwise}
    \end{cases}
    \label{eq:omegas_expression}
,\end{equation}

\noindent where $q \geqslant 1$ is the mass ratio between the dominant fragment and one satellite fragment. If only one child is produced, $n_l = 1$ and the mass ratio $q$ is by definition undefined. The Eq.~\ref{eq:omegas_expression} accounts for this case since whatever the value of $q$, the single child inherits all of the parental mass reservoir as $\psi_{l,i} = 1$. A uniform sibling mass distribution is modelled by $q = 1$ as all the children inherit the same fraction of the reservoir.

\subsection{Scale-free fragmentation}
\label{sec:Scale-free fragmentation}

The structure of the interstellar medium that constitutes a molecular cloud exhibits fractal properties \citep{elmegreen_falgarone1996} that reflect its self-similarity across different spatial scales. These self-similar properties supposedly arise from the scale-free nature of turbulence and gravity \citep{H&C2008, guszejnv2018_scalefree}. Although this self-similarity emerges within a limited range of spatial scales \citep{thomasson2024}, the cloud can always be described as locally scale-free. This scale-free approximation can be used to identify asymptotic and general trends regarding the effect of hierarchical fragmentation.

We introduce the fragmentation {spatial} rate $\phi$ \citep{thomasson2024} that {describes the rate at which the number of fragment grows}  between two successive spatial scales $R_{l+1}$ and $R_{l}$ such that

\begin{equation}
    \bar{n_l} = \left( \frac{R_{l+1}}{R_{l}} \right )^{-\phi}
    \label{eq:micron_phi}
.\end{equation}

Then, we introduce the mass transfer {spatial} rate $\xi$ \citep{thomasson2024} that {describes the rate at which the fragment mass efficiency grows between two successive spatial scales} as

\begin{equation}
    \epsilon_l = \left( \frac{R_{l+1}}{R_{l}} \right )^{-\xi}
    \label{eq:micro_epsilon_phi}
.\end{equation}

These relations ensure that for equal scale ratios $r = R_{l}/R_{l+1}$ the average number of fragments $\bar{n_l}$ produced and the mass efficiency $\epsilon_l$ are always the same. In a more general perspective, the choice of scale-free relations can be employed as an approximation in the case of slow variations of more complex functions $\phi(R)$ and $\xi(R)$ within sufficiently small spatial scale range such that $\phi$ and $\xi$ can be considered as constant. 

In this framework, \citet{thomasson2024} obtained the expression for the average mass $\langle M \rangle$ of individual fragments at any scale $R$ as

\begin{equation}
    \langle M_{l+1} \rangle = \langle M_{l} \rangle \left( \frac{R_{l+1}}{R_{l}} \right )^{\phi - \xi}
    \label{eq:average_mass_scale_free}
.\end{equation}

From Eq.~\ref{eq:average_mass_scale_free}, we understand that the average mass of a fragment not only depends on the mass efficiency through $\xi$, but also on the number of other fragments produced within the cloud through $\phi$. The more the cloud fragments, the higher the number of fragments is. The flipside of this production growth is that the fragments formed are less massive as they each have access to less material. In particular, clustered fragments tends to be less massive than single fragments for the same cloud total mass.

The scale-free nature of this model dictates that sub-fragmentation continues for infinitely small scales for $\phi > 0$. However, such feature is not physical since stars have to form as a result of a gravitational collapse. Thus, it is necessary in this scale-free model to set \textit{a posteriori} one stopping scale $R_\text{stop}$. This stopping scale sets the moment where a parental core does not sub-fragment anymore and gravitationally collapses into a single star, for example when the cores become opaque to their own radiation so non-thermal processes become dominant \citep{larson1969_fcore, Lee&H2018_adiafirstlarson} and additional fragmentation events are prevented \citep{thomasson2024}.

\subsection{Fragments random selection}
\label{sec:Fragments random selection}

The fragmentation rate $\phi$ determines the average number of children $\bar{n_l}$ produced by one parent according to Eq.~\ref{eq:micron_phi}. Therefore, any choice of probability distribution $p_l(n_l)$ must satisfy

\begin{equation}
    \bar{n_l} = \sum_{n_l} n_l p_l(n_l) = \left( \frac{R_{l+1}}{R_{l}} \right )^{-\phi}
    \label{eq:proba_distribution_phi}
,\end{equation}

\noindent where $n_l$ can take any positive value.

For the sake of simplicity we choose to consider in this work a binomial fragmentation process, according to the following rule: a parent generates either $1$ or $2$ children with respective probabilities $1 - p$ and $p$. This rule is employed in Sects.\ref{sec:Hierarchical fragmentation applied to a top-heavy CMF} and \ref{sec:Stellar clustering as a function of stellar mass} in order to investigate the detailed impact of stochastic fragmentation on the IMF and the resulting stellar systems. We exclude the case where $n_l = 0$ as non existing fragment does not count in the total mass distribution of fragments. developing Eq.~\ref{eq:proba_distribution_phi}, the probability $p$ can be expressed as a function of the fragmentation rate $\phi$ as

\begin{equation}
    p = r^{\phi} - 1
\label{eq:probability_phi}
,\end{equation}

\noindent where $r = R_{l}/R_{l+1} > 1$ is the scaling ratio between two successive levels. For example, to form on average $\bar{n_l} = 1.3$ fragments, a parent has a probability $1 - p = 0.7$ to fragment into $n = 1$ child and a probability $p = 0.3$ to fragment into $n = 2$ children. Based on this binomial probability distribution the expected value $\bar{n_l}$ satisfies $\bar{n_l} = r^{\phi}$ for any $r$. According to Eq.~\ref{eq:probability_phi} the choice of this binomial probability distribution remains valid as long as

\begin{equation}
    \phi < \dfrac{\ln(2)}{\ln(r)}
\label{eq:condition_r}
,\end{equation}

\noindent so that $p < 1$. Hereafter, we consider that as long as Eq.~\ref{eq:condition_r} is satisfied, fragmentation can occur so fragments can be formed using Eq.~\ref{eq:probability_phi}.

\subsection{Star formation efficiency}
\label{sec:Star formation efficiency}

In our model, any spatial scale $R_l$ may be considered as a potential scale $R_\text{stop}$ below which fragmentation stops. In the following, we assume that each of the last fragment populating the spatial scale $R_\text{stop}$ produces a single star. The CMF of those last fragments at $R_\text{stop}$ would coincide with the IMF if all their mass were used to form their star. We introduce an additional average efficency $\overline{\epsilon_*} = M_\text{tot}(R_*)/M_\text{tot}(R_\text{stop})$ that accounts for any star formation process below $R_\text{stop}$ that may influence the star formation efficiency. This star formation efficiency is used to connect the mass function of the last fragmenting cores with their stellar IMF. 

We define the net star formation mass efficiency $\mathcal{E}(R_*)$ for the cloud as the product of the cloud-to-fragments efficiency $\mathcal{E}(R_\text{stop})$ and the last fragments-to-star efficiency $\overline{\epsilon_*}$ as

\begin{equation}
    \mathcal{E}(R_*) = \mathcal{E}(R_\text{stop}) \times \overline{\epsilon_*}
    \label{eq:efficiencytotalsf}
,\end{equation}

\noindent where $\mathcal{E}(R_\text{stop}) = (R_\text{stop}/R_0)^{-\xi}$ from Eq.~\ref{eq:micro_epsilon_phi} provided that $\xi$ is scale-free.

\subsection{Local shape evolution}
\label{sec:Local shape evolution}

With this fragmentation framework, we evaluate the influence of hierarchical fragmentation on any mass distribution taking into account the amount of objects that appear in the cloud and their formation mass efficiency. To quantify how the shape of the CMF is affected by hierarchical fragmentation, we consider an average process in which all the fragments produced constitute a macroscopic ensemble, regardless of how siblings may share their parental mass reservoir among themselves. To determine the shape evolution of the CMF with fragmentation, we define as $\Gamma(R, M)$ the local logarithmic slope of the distribution at mass $M$ and spatial scale $R$ such that

\begin{equation}
    \dfrac{\partial N(R, M)}{\partial \log M}\Big|_R \propto M^{\Gamma}
    \label{eq:alpha_pdf}
,\end{equation}

\noindent where $\Gamma(R, M) = -1.35$ corresponds to the numerical value of the Salpeter slope for $M > 1 $M$_\odot$. We show in Appendix~\ref{appendix_demo_slopevar} that this local logarithmic slope $\Gamma(R, M)$ is determined by

\begin{equation}
    \dfrac{\partial \Gamma(R, M)}{\partial \log R} + \big(\phi - \xi) \dfrac{\partial \Gamma(R, M)}{\partial \log M} = \Gamma \dfrac{\partial \xi(M)}{\partial \log M} - (1 + \Gamma) \dfrac{\partial \phi(M)}{\partial \log M}
    \label{eq:final_slope_var}
.\end{equation}

This equation quantifies how the local logarithmic slope $\Gamma(R, M)$, at a mass bin $M$, varies with spatial scale $R$ during a hierarchical fragmentation process. Three terms contributes to vary this local slope. 

On the left-hand side of Eq.~\ref{eq:final_slope_var}, the advection term of parameter $\phi - \xi$ expresses the shift of the distribution along the mass domain. This advection term may reshape the distribution under some conditions described in Sect.~\ref{sec:Conditions to reshape a distribution}. Those shape variations are investigated in more details in Sects.\ref{sec:Retrieving the L3-cIMF from the fCMF} and \ref{sec:Stellar clustering as a function of stellar mass} in order to derive the cIMF from a top-heavy CMF along with the resulting stellar systems.

On the right-hand side of Eq.~\ref{eq:final_slope_var}, two mass derivative terms contribute to modify the local slope of the distribution. These terms account for the mass variations of both the fragmentation rate $\phi(M)$ and the mass transfer rate $\xi(M)$ and are analytically investigated in Sect.~\ref{sec:Impact of mass dependencies on a mass distribution}.

\section{Hierarchical fragmentation applied to a top-heavy CMF without mass dependencies}
\label{sec:Hierarchical fragmentation applied to a top-heavy CMF}

In the following we consider that both the fragmentation rate and the mass transfer rate are independent of the parental mass. Under such circumstance, only the influence of the advection term of Eq.~\ref{eq:final_slope_var} is adressed here. As an experiment, the initial CMF used to derive the fragmented CMF (fCMF) is parameterised by a power-law $dN/dlog M \propto M^{\Gamma}$ of index $\Gamma = -0.95$ ranging from 0.8$~{\rm M}_\odot$ to 100$~{\rm M}_\odot$. Such top-heavy CMF has been observed by \cite{pouteau2022} using the 1.3~mm and 3~mm ALMA 12~m array observation of the W43-MM2\&MM3 star forming region located at 5.5~kpc from our Sun \citep{zhang2014}. The angular resolution of these observations is associated to $R_0 = 2500$~AU spatial scale resolution below which we expect the cores to undergo scale-free fragmentation \citep{thomasson2024}. The slope value $\Gamma = -0.95 \pm 0.04$ has been measured with robust confidence above the completeness level $M$ {$\approx$} $0.8~{\rm M}_\odot$. We introduce our methodology to build the fCMF and compare it with the cIMF in Sect.~\ref{sec:Methodology to build the fCMF and compare it with the cIMF}. Then, in Sect.~\ref{sec:Retrieving the L3-cIMF from the fCMF}, we use this method to assess the fragmentation conditions that sufficiently reshape the initial distribution into a cIMF, and derive in Sect.~\ref{sec:Stellar systems formed through fragmentation} the resulting stellar clustering.

\subsection{{Consequences of the advection term}}
\label{sec:Consequences of the advection term}

\subsubsection{Shift of the initial distribution}

Considering a fragmentation process that do not depend on the parental mass, the Eq.~\ref{eq:final_slope_var} simplifies into an advection equation of parameter $\phi - \xi$:

\begin{equation}
    \dfrac{\partial \Gamma}{\partial \log R} + \big(\phi - \xi \big) \dfrac{\partial \Gamma}{\partial \log M} = 0
    \label{eq:final_slope_var_indep}
.\end{equation}

According to this advection equation, the logarithmic slope $\Gamma(R, M)$ {shifts} along both the mass domain and the spatial scales at a rate of $\phi - \xi$ {which} represents the mass efficiency of one individual fragment relative to its parent, considering the siblings share their parental mass reservoir (through $\phi$) and that this reservoir emerges with some efficiency (through $\xi$), as stated in Eq.~\ref{eq:average_mass_scale_free}. The Eq.~\ref{eq:final_slope_var_indep} remains valid for any subset of fragments that have been produced with the same individual mass efficiency. 

\subsubsection{Conditions to reshape a distribution}
\label{sec:Conditions to reshape a distribution}


Each mass bin $M_{l+1}$ constituting the distribution at a given scale $R_{l+1}$ is built from the contributions of all the parental masses $M$ within scale $R_{l}$ whose possible fragmentation outcomes may generate children of mass $M_{l+1}$. Thus, the local logarithmic slope $\Gamma(R_{l+1}, M_{l+1})$ at one child bin depends on the {local logarithmic slopes $\Gamma(R_{l}, M_{l})$ of} every possible parental mass bin.

If the children are produced with the same individual mass efficiency, then each child mass bin is constructed from a unique parental bin. Hence, the {$\Gamma(R_{l+1}, M_{l+1})$} of the different parental bins does not mix. Consequently, the distribution shifts to a different mass range and {its shape is not modified}. On the contrary, if at least two parental bins yield a fragmentation outcome that produces children in the same mass bin, the global shape can change. Therefore, {if parental cores produce fragments with different mass efficiencies (i.e. through $q \neq 1$), or if they have multiple outcomes for their number of fragments, the shape of the distribution can be modified.}

Hereafter, we consider the particular case of a power-law mass function whose logarithmic slope is constant along its mass range. Hence, $\partial \Gamma / \partial \log M = 0$ in Eq.~\ref{eq:final_slope_var_indep}. Regardless of how many outcomes can contribute to a child bin, its local slope $\Gamma(M_{l+1})$ remains similar with the slopes of its parental bins. For the distribution to be modified, the parental distribution must be bounded in mass. In that case, some of the fragmentation outcomes contributing to a child mass bin at the edges of the distribution may arise from unpopulated parental mass bins. On the other hand, more central child mass bins may be constructed from outcomes of populated parental bins. Then, the most extreme child bins inherit from a smaller population than their central counterparts, which effectively adds more objects to the central part of the child distribution compared to the edges, thus reshaping the distribution at its boundaries.

\subsection{{Method to compare the fCMF and the cIMF}}
\label{sec:Methodology to build the fCMF and compare it with the cIMF}

\subsubsection{Generation of a fCMF}
\label{sec:Generation of a fCMF}

In order to derive the fCMF, we introduce a semi-analytic procedural method. The following method remains valid assuming (i) $\epsilon_{l}$ is not a random variable and (ii) both the fragmentation rate $\phi$ and mass transfer rate $\xi$ do not depend on the mass of the parental object. The mass distribution of a population of fragments located inside any level $l$ is described by a PDF $\zeta_l(M)$ normalised as

\begin{equation}
    \displaystyle \int_{0}^{+\infty} \zeta_l(M) dM = 1
    \label{eq:Chap5:normzeta}
.\end{equation}

The mass function $\zeta_{l+1}(M)$ associated to the population of the next level can be derived from $\zeta_{l}(M)$ by considering every possible fragmentation outcome of the parents constituting $\zeta_{l}(M)$. These outcomes can be categorised with respect to the amount of children $n_l$ the parents may produce and the fraction of mass $\epsilon_l \psi_{l,i}(n_l)$ each individual children receive from their associated parent. For example, all the $i-th$ children originating from a $n_l = 2$ outcome who have received the fraction $\epsilon_l \psi_{l,i}(n_l = 2)$ from their parent, constitute one sub-population of $\zeta_{l+1}(M)$. This sub-population is associated with a specific individual mass efficiency quantified by $\epsilon_l \psi_{l,i}(n_l = 2)$. At level $l+1$, a sub-population is then characterised by the collection of fragments originating from the same pair ($n_l$ ; $\epsilon_l \psi_{l,i}$). We show in Appendix~\ref{app:Procedural generation of a fCMF} that considering every possible pair ($n_l$ ; $\epsilon_l \psi_{l,i}$) the fCMF at the next level can be written as

\begin{equation}
    \zeta_{l+1}(M) = \displaystyle \sum_{n_l} \sum_{i=1}^{n_l} \dfrac{p_l(n_l)}{\epsilon_l} \zeta_{l}\left(\dfrac{M}{\epsilon_l \psi_{l,i}(n_l)}\right)
    \label{eq:zeta_l_expression}
.\end{equation}

The construction of the mass functions of higher levels is procedural, considering each level one after the other {, starting from the initial CMF at $l = 0$.}


\subsubsection{Statistical comparison with the cIMF}
\label{sec:Statistical comparison with cIMF}

In order to compare our model's fCMF with the cIMF, we propose to use the log-logistic parametrisation by \cite{masch2013} designated as L3-cIMF. This is a non-segmented, smooth and continuous function described by three shape parameters $\gamma, ~\beta, ~\mu$ and two limits parameters $m_l, ~m_u$. The probability density $p_{L3}(m)$ of this functional is expressed as a function of $\gamma, ~\beta, ~\mu$

\begin{equation}
    p_{L3}(m) = G(m_l, m_u) \left( \frac{m}{\mu} \right)^{-\gamma} \Biggl( 1 +  \left( \frac{m}{\mu} \right)^{1-\gamma} \Biggr)^{-\beta}
.\end{equation}

With $G(m_l, m_u)$ a normalisation coefficient. The limits $m_l, m_u$ represent respectively the lower and upper limit masses of the probability distribution $p_{L3}$. The canonical parameters that describe the cIMF are given by $\gamma = 2.3$, $\beta = 1.4$ and $\mu = 0.2~{\rm M}_\odot$. The parameter $\gamma$ represents the logarithmic slope at large masses equivalent to the Salpeter slope. The parameter $\beta$ is related to the logarithmic slope at low-masses while the parameter $\mu$ is caracteristic of the peak of the cIMF. We set the range of masses starting from $m_l = 0.01~{\rm M}_\odot$ onward to $m_u = 150~{\rm M}_\odot$. 

This cIMF is compared with our fCMF using the Anderson-Darling (AD) test \citep{andersondarling1954} that compares two distributions. This test is a non-parametric statistical test based on the empirical cumulative distribution {that has the advantage of being equally sensitive over the whole sample range.} Let $H_0$ be the null hypothesis stating that a sample of $N$ fragments selected from the fCMF may in fact come from the L3-cIMF distribution. The statistic $A^2$ of this test is computed as follow:

\begin{equation}
    A^2 = - N - \frac{1}{N} \displaystyle\sum_{i=1}^N (2i - 1) \Bigl[\ln( F(m_i) ) + \ln(1 - F(m_{N-i+1})\Bigr]
.\end{equation}

Where $F(m)$ is the cumulative distribution function of the L3 cIMF, and $m_i$ is the $i-th$ element of the vector $m$ that represents the ordered masses from the smallest to the largest value within a given sample. We perform the AD test using the generic critical values with their associated significance levels (Table \ref{tab:score_test}). A fCMF is considered compatible with the cIMF if we cannot reject the null hypothesis at the significance level of 0.05, that is when the p-value is $>$ 0.05.

\begin{table}[]
    \centering
    \caption{Statistics of the AD test associated with their significance level from \cite{dagostino1987}. We checked the validity of these values by reconstructing the distribution of the AD statistics from the L3-cIMF of \cite{masch2013} with a bootstrap technique.}
    \begin{tabular}{l  c c c c c c c c c}
    \hline \hline
    p-value & 0.1 & 0.05 & 0.025 & 0.01 & 0.005 & 0.001 \\
    $A^2$ & 1.933 & 2.492 & 3.070 & 3.880 & 4.500 & 6.000\\
    \hline
    \end{tabular}
    \label{tab:score_test}
\end{table}

%

\subsection{Retrieving the L3-cIMF from the fCMF}
\label{sec:Retrieving the L3-cIMF from the fCMF}

\subsubsection{Model setup}
\label{sec:Model setup}

The L3-cIMF is compared with the fCMFs derived from different sets of parameters $\{\phi, \xi, q \}$, respectively the fragmentation rate, the mass transfer rate and the fragment mass ratio. We monitor the evolution of the fCMF across 8 levels of fragmentation from $R_0 = 2500$~AU to $R_8 \approx 98$~AU. We vary $\phi$ between 0 and 1.5, corresponding to a fragmentation probability $0 < p < 0.84$ between two successive scales ; $\xi$ between -1 and 0.5, corresponding to efficiencies $67\% < \epsilon_l < 122\%$, meaning that mass accretion is possible. Each level is separated by a scaling ratio $r = 1.5$ in order to model a binary fragmentation between each level (see Sect.~\ref{sec:Fragments random selection}). Since we consider a scale-free process, $r$ only regulates the number of possible fragmentation events between the initial scale $R_0$ and the final scale $R_\text{stop}$. The fragments mass ratio $q$ vary between 1 and 5. We sample the obtained fCMFs with $N = 1,000$ objects in order to test each fCMF against the L3-cIMF from $M = 10^{-2}$M$_\odot$ up to $M = 10^{2}$M$_\odot$ using the AD test. We show in Fig.~\ref{fig:space_solution_example} diagrams displaying the solutions with which $H_0$ cannot be rejected at a 0.05 significance level.

\subsubsection{{Convergence to a unique shape}}
\label{sec:Degeneracy of fCMF solutions}

\begin{figure*}[!t]
    \centering
    \includegraphics[width=18cm]{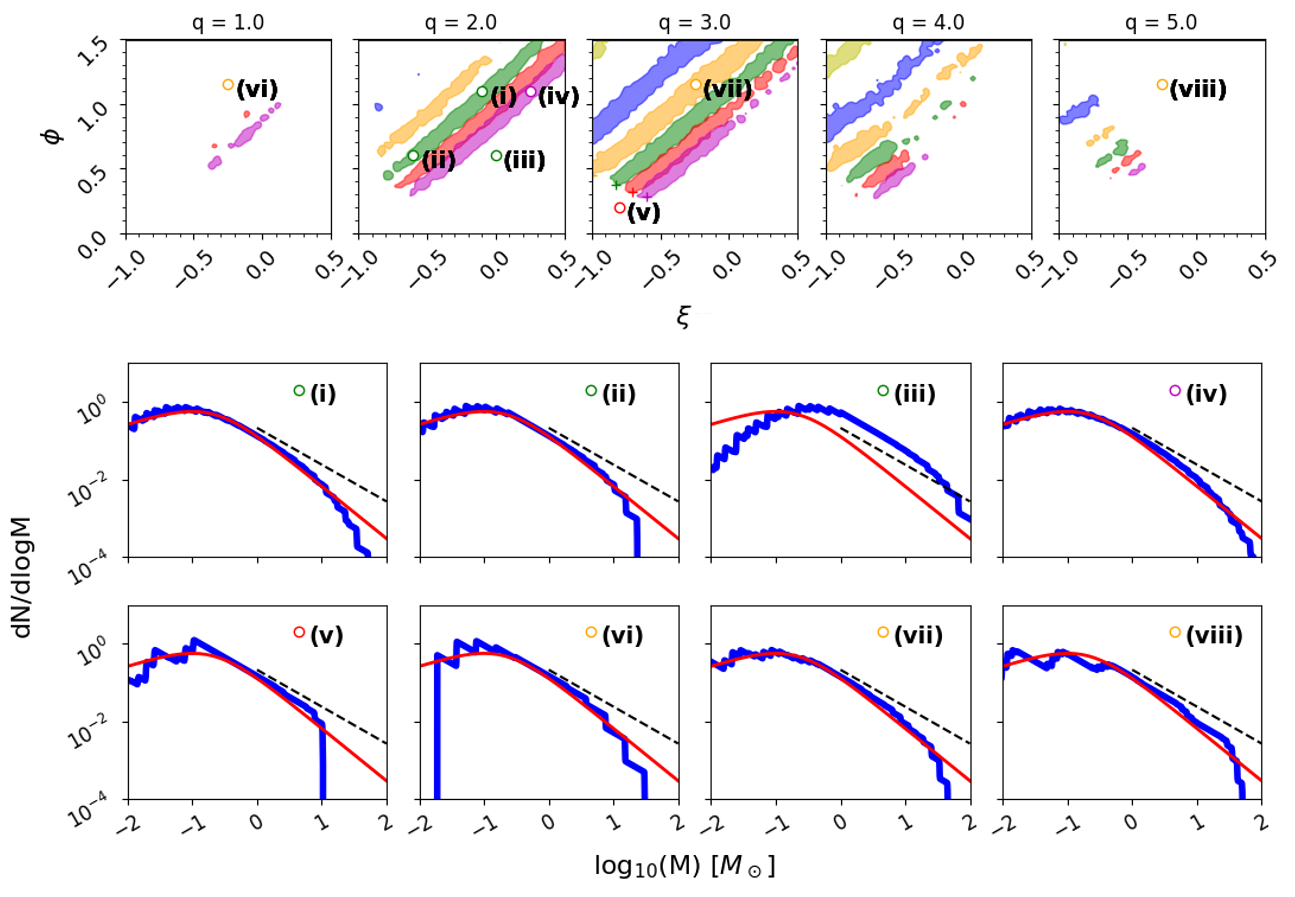}
    \caption{Parameter space for which a fCMF is compatible with the \cite{masch2013} cIMF in the sense of the AD test. \textit{Top}: Parameter space at which a fCMF is compatible with the cIMF at 0.05 confidence level, in the fragmentation rate ($\phi$) vs mass transfer rate ($\xi$) diagram for different fragments mass ratio $q$. Yellow, blue, orange, green, red and magenta patches highlight the solutions at $R_\text{stop} = 741, 494, 329, 219, 146, 98$~AU after $l = 3, 4, 5, 6, 7, 8$ levels of fragmentation, respectively. Crosses at $q = 3.0$ indicate points where the average total number of fragments produced at $R_\text{stop}$ is {$\approx$}2.5. \textit{Bottom}: fCMF associated to the coloured-labelled points. Blue, red and black dashed lines represent the fCMF, cIMF and the slope of the initial top-heavy CMF respectively. The AD test is performed from $M = 0.2$~M$_\odot$ to $M = 150$~M$_\odot$ to compare the fCMF with the cIMF.}
    \label{fig:space_solution_example}
\end{figure*}

{The AD test shows that some fCMFs are similar to the L3-cIMF (Fig.\ref{fig:space_solution_example}) when $\phi$ and $\xi$ are linearly correlated, assuming a final fragments-to-star efficiency of $\overline{\epsilon_*} = 100\%$. This degeneracy is a consequence of Eq.~\ref{eq:average_mass_scale_free}, as the average total mass that a parent transfers to its children depends on both $\xi$ and $\phi$, such that $\phi - \xi$ needs to be constant to obtain the same average mass, for the same spatial scale. On the other hand, the mass ranges of the fCMF and L3-cIMF must coincide in order for the fCMFs to be compatible with the L3-IMF. Thus, for each spatial scale, all valid fCMF solutions are shifted by the same amount in mass. For example, panels (i) and (ii) in Fig.~\ref{fig:space_solution_example} are valid cIMF solutions, whereas panel (iii) is not. \\ This convergence also depends on the spatial scale $R_l$. As $R_l$ decreases, solutions shift towards lower fragmentation rates and higher mass transfer rates (see e.g.(i)-(iv) of Fig.~\ref{fig:space_solution_example}). Producing fragments of same mass at smaller scales requires keeping more efficiently the fragment mass between two successive scales, which is achieved by fragmenting less, and transfering more mass into the children. At a given stopping scale $R_\text{stop}$, this degeneracy breaks down when the average number of fragments falls below $\langle N_\text{tot}(R_\text{stop}) \rangle \lesssim 2.5$. Beyond that point, the resulting fCMF is indistinguishable from a scenario without fragmentation, yielding a single outcome in which one fragment is likely produced (see e.g. (v) of Fig.~\ref{fig:space_solution_example}). \\ Although the fCMFs emerge from different sets of parameters ($\phi$, $\xi$, $q$, $R_\text{stop}$), the resulting fCMFs exhibit similar shapes, resembling the L3-IMF. For example, panels (ii) and (iii) have different $\xi$ and mass range, but share the same shape. This implies that for any fragmentation setup, one can always find a value of $\overline{\epsilon_*}$ that reconciles the mass ranges of the fCMF with that of the cIMF. The W43-MM2\&3 top-heavy CMF converges toward similar shape under mass-independent hierarchical fragmentation, suggesting that such a process yields a universal outcome, at least within the explored parameter space.}

\subsubsection{{Limits of convergence}}
\label{sec:Retrieving the shape of the cIMF}

{We recall that we assumed no dependence of $\phi$ and $\xi$ on the initial core mass. Under this assumption, a power-law CMF is reshaped if it has mass boundaries (Sect.\ref{sec:Conditions to reshape a distribution}), allowing variations at the distribution edges to propagate inward. In real molecular clouds, finite core samples naturally impose such boundaries, and observed CMFs rarely exhibit perfect power-law shapes due to statistical uncertainties \citep[e.g.][]{enoch2007, konyves2020, Ladjelate2020}. In our case, the CMF is reshaped through the probabilistic number of fragments formed and the non-uniform mass partitioning, considering both equal-mass fragmentation ($q=1$) and unbalanced cases ($q > 1$). A L3-IMF solution only arises for mass ratios $2 \lesssim q \lesssim 4$ when there are fewer than seven fragmentation levels ($l<7$). When $q=1$, mass-independent fragmentation alone is insufficient to reshape the CMF into the L3-IMF (see (vi)-(vii) of Fig.~\ref{fig:space_solution_example}). Thus, the mass ratio is important to efficiently modify the CMF within limited fragmentation levels. \\ At low masses, different fragmentation outcomes produce a log-normal cut-off, forming the IMF peak. At high masses, the distribution steepens toward the Salpeter IMF up to $\approx 10$~M$_\odot$, beyond which the distribution diverges from the Salpeter slope. Such high-mass cut-off is expected from fragmentation when stochasticity is introduced \citep{larson1973_fragmentation}. In analytical theories of CMF arising from isothermal, turbulent, and/or thermally supported clouds, the location of such cut-off at high-mass depends on the cloud Mach number \citep{H&C2008}. In addition, cloud sub-fragmentation appears to lower the characteristic mass of this cut-off \citep{hopkins2012_last_crossing}, as structures become comparatively less massive. However, power-law behavior still dominates for masses $M < 100~$M$_\odot$, when the cloud Mach number $\mathcal{M} > 6$.} 


\subsubsection{Beyond Core Fragmentation}

{Our study focuses on the fragmentation of compact prestellar cores. The framework we develop is independent of any specific fragmentation mechanism. The fragmentation rate $\phi$ and mass transfer rate $\xi$ are treated as free parameters, and the stochasticity reflects our limited knowledge of how mass is redistributed during fragmentation. Our framework remains applicable to various star-forming environments, including non-turbulent filaments \citep{andre2010, hacar2017, andre2019}. Starting from an initial filament mass function \citep[e.g.][]{andre2019}, we can test which combinations of $\phi$ and $\xi$ reproduce the observed IMF or CMF, and identify the fragmentation conditions required to link the initial filamentary structure to the final stellar population. Then, these conditions can be tested against empirical measurements \citep{thomasson2024} or using numerical simulations.}

\subsubsection{{Brown dwarfs formation}}

The stellar part of the IMF declines sharply around $0.2~M_\odot$, where the brown dwarf (BD) population emerges. Thus, the low-mass end of the cIMF likely consists of two overlapping distributions \citep{thies2015}. In our comparison between fCMFs and the cIMF, we did not explicitly account for the contribution of BDs, which may produce a bimodal IMF \citep{Drass2016}. We assumed that all objects follow a similar formation process to the stars as supported by recent observations of proto- and pre-BD cores \citep{Palau2024}. Thus, BD-cores undergo the same scale-free hierarchical fragmentation as pre-stellar cores.
Discrepancy between the L3-IMF we used and the BDs IMF suggest that BDs form through alternative channels \citep[see][]{kroupabouvier2003, Whitworth2018}. It can also means they go through different fragmentation regimes compared to stars, possibly involving non-scale-free behavior, mass-dependent processes, or stronger environmental influence.


\section{Impact of mass dependencies on a mass distribution}
\label{sec:Impact of mass dependencies on a mass distribution}

In Sect.~\ref{sec:Hierarchical fragmentation applied to a top-heavy CMF}, we {assume the fragmentation rate and the mass transfer rate do not depend on the core mass}. In fact, the chosen fragmentation prescription may depend on the local physical properties of the cloud as every fragmentation theories involve density instabilities for a gas structure to gravitationally collapse \citep{jeans1902, hopkins2012_last_crossing}. In the following, we investigate the impact of mass dependencies of the fragmentation rate $\phi(M)$ and mass transfer rate $\xi(M)$ on the shape of the fCMF, quantified by the right-hand side of Eq.~\ref{eq:final_slope_var}, so we assume $\frac{\partial \Gamma}{\partial log M} = 0$.


\subsection{{Global solution}}


{To quantify the variation of the local slope $\Gamma(R, M)$ under mass dependent fragmentation processes only,} we rewrite Eq.~\ref{eq:final_slope_var} ignoring the advective term

\begin{equation}
    \dfrac{\partial \Gamma}{\partial \log R} = \Gamma \dfrac{\partial \xi}{\partial \log M} - (1 + \Gamma) \dfrac{\partial \phi}{\partial \log M}
.\end{equation}

\noindent {Hereafter, we let $\xi'_M = \partial \xi / \partial \log M$ and $\phi'_M = \partial \phi / \partial \log M$}. Assuming constant $\xi'_M$ and $\phi'_M$ to assess asymptotic trend, the solution to this differential equation is

\begin{equation}
    \Gamma(R, M) = \dfrac{\phi'_M}{\xi'_M - \phi'_M} + \left(\dfrac{R}{R_0}\right)^{\xi'_M - \phi'_M} \left[ \Gamma_0(M)  - \dfrac{\phi'_M}{\xi'_M - \phi'_M} \right]
    \label{eq:alpha_variation_phi_xi}
,\end{equation}

\noindent where $\Gamma_0$ is the initial logarithmic slope of the CMF at scale $R_0$. 


\subsection{Effect of the mass transfer rate variations $\xi(M)$}
\label{sec:Effect of the mass transfer rate}

To investigate the impact of the mass transfer rate $\xi(M)$ {only}, we {set $\phi'_M = 0$ in Eq.\ref{eq:alpha_variation_phi_xi}, that becomes}

\begin{equation}
    \Gamma(R, M) = \Gamma_0(M) \left( \dfrac{R}{R_0} \right)^{\xi'_M}
    \label{eq:alpha_variation_xi}
,\end{equation}

If more massive objects form their fragments more efficiently in terms of mass as they are more gravitationally bound to their environment, so we can write $\xi'_M > 0$. The slope converges to $\Gamma(M) = 0$ as hierarchical fragmentation proceed (as $R$ decreases). Therefore, if $\xi'_M > 0$, $\Gamma = 0$ appears as an asymptotic attractor for any initial value $\Gamma_0(M)$. On the contrary, if ${\xi'_M} < 0$, the less massive cores transfer their mass more efficiently to their children. As expressed by Eq.~\ref{eq:alpha_variation_xi}, $\Gamma$ diverges in this case.

To complement, we checked these theoretical variation trends in Appendix~\ref{app:Mass dependent model validity} using a Monte-Carlo simulations that provide good agreement with Eq.~\ref{eq:alpha_variation_xi}.

\subsection{Effect of the fragmentation rate variations $\phi(M)$}
\label{sec:Effect of the fragmentation rate}

To investigate the impact of the fragmentation transfer rate $\phi(M)$ {only}, we {set $\xi'_M = 0$ in Eq.\ref{eq:alpha_variation_phi_xi}, that becomes}

\begin{equation}
    \Gamma(R, M) = \big[ \Gamma_0(M) + 1 \big] \left( \dfrac{R}{R_0} \right)^{-\phi'_M} - 1
    \label{eq:alpha_variation_phi}
,\end{equation}

If $\phi'_M < 0$, the slope converges to $\Gamma(M) = -1$ as hierarchical fragmentation{continues at smaller scales}. $\Gamma = -1$ appears as an asymptotic attractor for any initial value $\Gamma_0(M)$. On the contrary, if ${\phi'_M} > 0$, $\Gamma$ diverges from the asymptotic value $\Gamma = -1$. 

We checked these theoretical variations in Appendix~\ref{app:Mass dependent model validity} using Monte-Carlo simulations that provide good agreement with Eq.~\ref{eq:alpha_variation_phi}.

\subsection{Convergence to the Salpeter IMF}
\label{sec:Convergence to the Salpeter IMF}

\begin{figure}
    \centering
    \includegraphics[width=9cm]{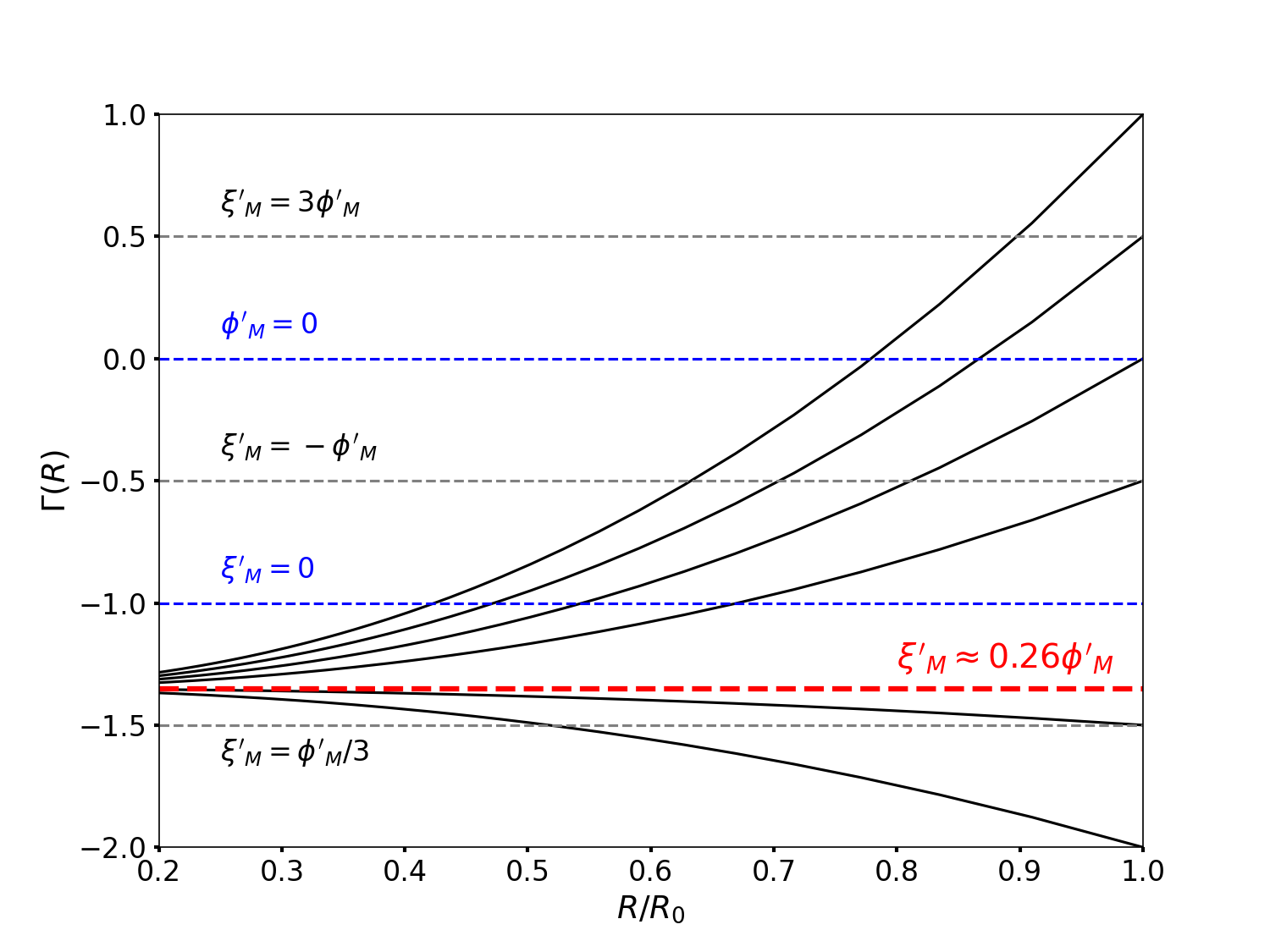}
    \caption{Convergence of the local logarithmic slope $\Gamma(R, M)$ under the influence of hierarchical fragmentation with the spatial scales $R$ (solid black line) towards the Salpeter slope (dotted red line) for different initial slopes $\Gamma_0 = -2.0, -1.5, -0.5, 0, 0.5, 1.0$. This convergence is only possible if $\xi'_M > \phi'_M$, with $\xi'_M = \partial \xi / \partial \log M$ and $\phi'_M = \partial \phi / \partial \log M$. The initial distribution can also converge towards other asymptotic values (dotted grey lines) depending on the relationship between $\xi'_M$ and $\phi'_M$ according to Eq.~\ref{eq:alpha_variation_phi_xi}. The slopes associated with the cases $\xi'_M = 0$ and $\phi'_M = 0$ (dotted blue lines) correspond to the asymptotes determined in Sects.\ref{sec:Effect of the mass transfer rate} and \ref{sec:Effect of the fragmentation rate} respectively.}
    \label{fig:convergence_slope}
\end{figure}

{In the Sects.\ref{sec:Effect of the mass transfer rate} and \ref{sec:Effect of the fragmentation rate}, we decoupled the individual effects of $\xi(M)$ and $\phi(M)$. We now identify more general solutions that lead a CMF to converge towards a power-law distribution. We consider the coupled influence of $\xi(M)$ and $\phi(M)$.} We recall that in our model, as fragmentation occurs, the spatial scale $R$ decreases. Hereafter we focus on the asymptotic conditions $R \ll R_0$ in which the fragmentation processes are finished. Under these conditions, $\Gamma(R) \equiv \Gamma_\text{IMF}$ the local slope of the IMF at a given mass $M$. In addition, Eq.~\ref{eq:alpha_variation_phi_xi} converges at low scales when $\xi'_M - \phi'_M > 0$. Its asymptotic limit reads

\begin{equation}
    \xi'_M = \dfrac{\Gamma_\text{IMF} + 1}{\Gamma_\text{IMF}} \phi'_M
    \label{eq:asymptotic_stable_IMFsolution}
.\end{equation}

If we consider that the slope of the IMF corresponds to the Salpeter slope, $\Gamma_\text{IMF} = -1.35$ so we obtain $\xi'_M \approx 0.26 \phi'_M$. Hence, both the condition $\xi'_M - \phi'_M > 0$ and Eq.~\ref{eq:asymptotic_stable_IMFsolution} can be satisfied if $\phi'_M < 0$ which also implies $\xi'_M < 0$. 

Since the convergence to the Salpeter slope is independent of the initial slope $\Gamma_0$, every segment of the initial CMF have the same convergence regardless of the local variations of $\Gamma_0(M)$ along the initial mass domain {(i.e. regardless of the CMF shape).}

The condition $\phi'_M < 0$ corresponds to a fragmentation scenario in which the more massive a core, the less it fragments. Although counter-intuitive, this perspective may be possible as the density structure of more massive objects are more strongly determined by gravity. The density profile of these structures may tend to be statistically more radially concentrated \citep{shu_self_similar_1977}, preventing more easily the emergence of density enhancements that may grow to become unstable, for example due to turbulence \citep{kritsuk2011_rhopdf}. In these conditions, more massive structures tend to sub-fragment less compared to less massive structures. 

Because of gravity, we expect that more massive object bound their surrounding material more easily. This intuitive result suggests that more massive objects fragment with a higher mass efficiency, either because of higher infall rate \citep{Yue2021} or because their fragments lose less material from the parental reservoir \citep{louvet2014}. However, the condition $\xi'_M < 0$ corresponds to a scenario in which more massive objects are less efficient in producing their fragments. This apparent opposition is a matter of definition. In our model, the mass transfer rate $\xi$ is a parameter that tracks the mass efficiency (Eq.~\ref{eq:micro_epsilon_phi}). This formation efficiency is defined as the ratio between the mass of children on scale $R_{l+1}$ and the mass of their respective parents on scale $R_l$. In absolute value, the mass accretion rate may be higher for more massive fragments, but the proportion of accreted mass with respect to the parental mass can be lower (i.e. the mass efficiency can be lower for more massive fragments, so $\xi'_M < 0$). 

If a universal Salpeter IMF does indeed emerge from the CMF via hierarchical fragmentation processes, both conditions $\phi'_M < 0$ and $\xi'_M < 0$ {have to} remain valid for the most massive objects, at least during the end of the star formation process. In this context of hierarchical fragmentation, the convergence towards the classical Salpeter IMF requires specific conditions. The processes regulating both the fragments formation efficiency and the fragmentation {have to} balance with $\xi'_M \approx 0.26 \phi'_M$ and both $\xi'_M$ and $\phi'_M < 0$. 

As long as the convergence condition $\xi'_M > \phi'_M$ is satisfied, a stable power-law distribution can naturally emerge. However, the logarithmic slope associated to this stable power-law may not correspond to the Salpeter IMF if the prefactor connecting both parameters $\xi'_M$ and $\phi'_M$ is different than {$\approx$}0.26. Hence, the convergence of any CMF towards a Salpeter IMF through hierarchical fragmentation remains highly circumstantial and theoretical without additional contraints on this prefactor. Outside of these convergence conditions, if $\xi'_M < \phi'_M$ a Salpeter IMF may still emerge from a CMF coincidentally at some stopping scale $R_\text{stop}$ (Eq.\ref{eq:asymptotic_stable_IMFsolution} and Fig.\ref{fig:convergence_slope}).

%
%
\section{{Stellar systems formed through fragmentation}}
\label{sec:Stellar systems formed through fragmentation}


\label{sec:Stellar clustering as a function of stellar mass}

 
{To build the stellar systems resulting to fragmentation and compute their properties, we} perform $10^4$ random fragmentation draws of the sample constituting the W43-MM2\&MM3 CMF \citep{pouteau2022}, with $\phi = 1.0$, $\xi = -0.1$, $q = 2$ down to the scale $R_\text{stop} = 219$~AU, as we expect this solution to correspond to the cIMF (Sect.~\ref{sec:Degeneracy of fCMF solutions}). 


\begin{figure}
    \centering
    \includegraphics[width=9cm]{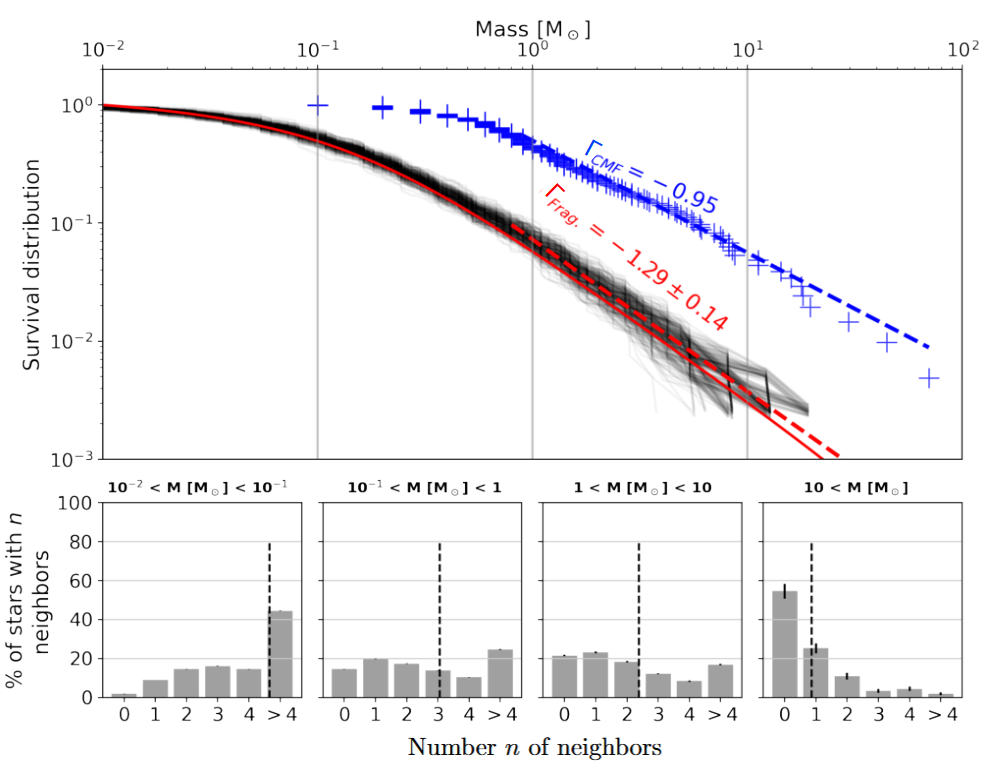}
    \caption{Survival function of the fragmented CMF correlated with the multiplicity of stellar systems formed at different mass interval. \textit{Top}: We perform $10^4$ random fragmentation draws (black lines) from the sample constituting the W43-MM2\&MM3 CMF as presented in \citet{pouteau2022} (blue crosses), using parameters $\phi = 1.0$, $\xi = -0.1$, and $q = 2$. Fragmentation covers spatial scales from $R_0 = 2500$~AU down to $R_\text{stop} = 219$~AU. For visibility, only the first 100 Monte Carlo draws are shown. The solid red and dashed blue line represents the canonical IMF and the power-law distribution fit from \citet{pouteau2022} respectively. The dashed red line represents the average of the slopes obtained from fitting each Monte Carlo draw, using masses $M > 0.8~$M$_\odot$. \textit{Bottom}: Distribution of the number of neighbors that a star of mass $M$ possesses in different mass interval indicated above each plot. The mean value is indicated by the horizontal dashed line.}
    \label{fig:mult_vs_mass}
\end{figure}

\subsection{{Stellar systems clustering}}
\label{sec:Stellar systems multiplicity}

By definition, the higher the fragmentation rate, the greater the average number of stars. At constant mass efficiency, the mass of a fragment varies inversely with its number of siblings ({as} $1/n$, Eqs.~\ref{eq:mass_of_children} and \ref{eq:omegas_expression}), which is a random variable. So, within a full population of fragments, the ones with the lowest masses are also those that have fragmented the most and are part of systems with the highest stellar density. In that case, stellar density refers to the number of stars located within a parental core of size $R_0 = 2500$~AU. Hence, the most massive stars that result from non-fragmented outcomes are born in isolation. For example, stars of mass $M < 0.1~$M$_\odot$ possess, on average, 4.6 neighbors in a $R_\text{core} = 2500$~AU vicinity, while stars of mass $M > 10~$M$_\odot$ possess 0.9 neighbors on average within the same vicinity (Fig.~\ref{fig:mult_vs_mass}). Since the average number of fragments produced by each parent depends on the fragmentation rate, these multiplicity values also depends on the fragmentation rate. Nonetheless, the tendency to produce multiple systems of high multiplicity order with low-mass stars remains for different fragmentation rate values. Although hierarchical fragmentation inherently favors the formation of multiple systems, the more a system contains stars, the lower their mass. Yet, these multiple systems are dynamically unstable so we expect them to quickly decay into multiple isolated stars. On the other hand, most massive objects that tends to be formed in isolation or in binaries are more stable. 

\subsection{{Multiplicity fractions through fragmentation}}
\label{sec:Stellar systems multiplicity fractions}

\begin{figure}
    \centering
    \includegraphics[width=9cm]{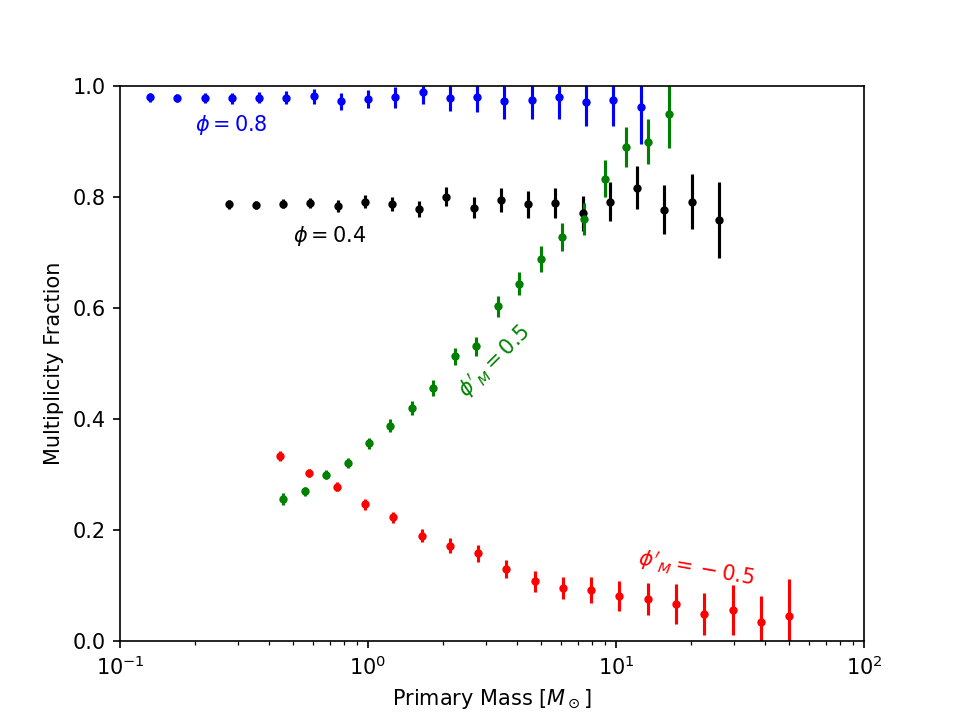}
    \caption{{Multiplicity fraction as a function of the primary masses grouped inside bins, for different fragmentation scenarios. Each Monte-Carlo simulation is carried out using $10^4$ random cores drawn from a top-heavy CMF with $\Gamma = -0.95$ \citep{pouteau2022}, with $\xi = -0.2$, $q = 2$ and 8 levels of fragmentation. Blue and black, green and red dots represent simulations using $\phi = 0.8, \phi = 0.4, \phi'_M = 0.5, \phi'_M = -0.5$.}}
    \label{fig:mf_vs_pm}
\end{figure}

{The multiplicity fraction (MF) represents the fraction of stellar systems whose primary star possesses at least one companion, with a specific separation. We introduce here the fragmented multiplicity fraction (fMF) as the MF resulting from the sub-fragmentation of a parental core, with separations $< R_0$. We explore here the qualitative trend of the fMF, as a function of fragmentation rate mass gradient $\phi'_M$. We can estimate a fragmented MF (fMF) as the probability that at least two stars are formed after the fragmentation process of a core. The lower the stopping scale $R_\text{stop}$, the higher the fMF for each primary mass, as the parental core has more opportunities to fragment. Similarly, the higher the fragmentation rate, the higher the fMF because the probability to fragment at each level increases. \\ 
When the fragmentation rate is independent of the core mass ($\phi'_M = 0$), the resulting fMF is constant with mass (see blue and black dots in Fig.\ref{fig:mf_vs_pm}). It reaches 100\% when the fragmentation probability approaches 100\%. If the gradient is positive ($\phi'_M > 0$), higher-mass cores fragment more, and the fMF increases for more massive primaries (see green dots in Fig.\ref{fig:mf_vs_pm}). If the gradient is negative ($\phi'_M < 0$) , higher-mass cores fragment less, and the fMF decreases for more massive primaries  (see red dots in Fig.\ref{fig:mf_vs_pm}).}

\section{Discussion}
\label{sec:discussion}

\subsection{High-mass part of the IMF}
\label{sec:High-mass part of the IMF}

The high-mass tail of the IMF is determined as a balance between the mass gradients of the fragmentation and the mass transfer rates (see \ref{eq:asymptotic_stable_IMFsolution}). We showed in Sect.\ref{sec:Convergence to the Salpeter IMF} that a Salpeter-like IMF arises when $\phi'_M < 0$, $\xi'_M < 0$, and $\xi'_M/\phi'_M \approx 0.26$, regardless of the shape of the initial CMF. This indicates that the fragments production efficiency increases roughly four times faster than the fragments mass efficiency for decreasing core mass. Top-heavy IMFs can be obtained by increasing the ratio $\xi'_M/\phi'_M$, i.e., by increasing $\xi'_M$ or decreasing $\phi'_M$. Such variations imply that top-heavy IMF are formed through different fragmentation properties, resulting from different environmental conditions \citep[e.g.]{Haslbauer2024, Zhang2018}. 

More generally, we expect the high-mass slope of the IMF to vary if the ratio $\xi'_M/\phi'_M$ fluctuates. However, if the IMF slope deviates by 10\% from the Salpeter value of $\Gamma = -1.35$, we expect the ratio to vary by 29\% (Eq.\ref{eq:asymptotic_stable_IMFsolution}). Consequently, the high-mass slope $\Gamma$ does not appear to be highly sensitive to fluctuations in fragmentation properties. This result provides a way to test our model by comparing the fragmentation properties of regions exhibiting top-heavy, and Salpeter IMFs. Moreover, further investigation is needed to identify the physical mechanisms determining the approximative value of 0.26 for the ratio $\xi'_M/\phi'_M$ that is necessary to obtain a Salpeter slope in the IMF.

\subsection{Multiplicity fraction}
\label{sec:Multiplicity fraction}

As shown in Sect.~\ref{sec:Stellar systems multiplicity fractions}, having a negative gradient $\phi'_M$, as requested to obtain a Salpeter-like IMF, yields to multiplicity fractions that appear inconsistent with observations. In our fCMFs, the most massive mass bins are populated by fragmentation outcomes associated with a single fragment produced ($n_l = 1$), meaning that the most massive stars are those with the lowest fMF (Fig.~\ref{fig:mult_vs_mass}). However, the observed proportion of multiples compared to isolated stars tends to increase with the mass of the primary object contained in stellar systems \citep{guzs2017_multiplicity, offner2023}. This suggests that the fragmentation rate $\phi$ should be high for high-mass cores and that $\phi'_M\geq 0$. {These observational constraints have been obtained for main-sequence stars, and it has been argued that primordial multiplicity fractions could be equally high for all primary masses (\citealt{kroupa2024} and references therein). In that case, we expect mass-independency of the fragmentation rate $\phi'_M = 0$. Measuring the fragmentation rate for different core masses is needed to discriminate the two possibilities.}

If the fragmentation rate increases significantly with the mass of the initial objects, \cite{houghton2024} found that with $\Gamma_0 > -1$, the slope of the fCMF becomes too steep compared to the Salpeter slope. This steepening effect is consistent with our fragmentation model, which predicts a steepening of $\Gamma$ if $\phi'_M > 0$. However, if $\phi(M)$ slowly varies with the mass ($\phi'_M \approx 0$), \cite{houghton2024} also showed that the slope variation is small enough to maintain the Salpeter slope while accounting for the MF of massive and low-mass primaries. 

\subsection{Massive star formation}
\label{sec:Massive star formation}

If the fragmentation rate is high for high-mass cores and $\phi'_M\geq 0$ as suggested above, the likelihood of cores forming massive stars decreases, thus resulting in a lower fraction of massive stars ($M > 10$~M$_\odot$) in our fCMFs compared to the Salpeter IMF (see fCMFs Fig.~\ref{fig:space_solution_example}). This depletion at high-masses can also be predicted in theoretical CMFs derived from isothermal turbulence \citep{H&C2008, hopkins2012_last_crossing, hopkins2013fragtheory}, or from other probabilistic models \citep{elmegreen1983, zinnecker1984}. Within the framework of our model, we interpret this depletion as a consequence of the stochastic nature of our fragmentation prescription, which naturally creates a log-normal distribution \citep{larson1973_fragmentation}, influenced by the mass ratio among the fragments formed (see Sect.~\ref{sec:Conditions to reshape a distribution} for conditions that affect the distribution shape).

In order to obtain a power-law behavior at high masses, as suggested by CMF from both cloud observations \citep{nutter2007, sokol2019, caoY2021, pouteau2022} and simulations of isothermal clouds undergoing gravo-turbulent fragmentation \citep{schmidt2010, Guszejnov2015Mapping}, and to get a number of high-mass stars ($M > 10$~M$_\odot$) consistent with the cIMF, $\xi'_M$ needs to be positive. This suggests that the core mass is positively correlated with the formation efficiency \citep{louvet2014}. Such a gradient in mass effectively flattens the stellar IMF locally and increases the number of massive stars compared to lower mass stars. The fCMF then consists of a low-mass log-normal shaped population resulting from stochastic fragmentation, and a high-mass power-law shaped population \citep{basu2004}, produced by a positive gradient in mass efficiency. This scenario is consistent with $\approx 0.1~$pc CMF observed in the Cygnus-X molecular cloud \citep{guangxingli2021}, which highlights a transition at $M \approx 10~$M$_\odot$ between a log normal and a power-law mass functions composed of fragmented and accreting cores respectively.

Moreover, having $\xi'_M>0$ also leads to the formation of primary objects with a higher fMF in existing stellar systems, as clustered low-mass stars may collectively accrete material from their environment while competing for it \citep{clark2021}. In this scenario, the probability of forming clustered stars with a massive primary increases, so the fMF becomes qualitatively more consistent with observations (see Sect.~\ref{sec:Multiplicity fraction}).

\subsection{Limitation of scale-free hierarchical fragmentation}
\label{Limitation of scale-free hierarchical fragmentation}

{We discuss the conditions that set both the turn-over mass of the cIMF with hierarchical fragmentation and the power-law at high mass. We distinguish three cases.
When mass gradients remain constant and exceed stochastic contributions (Sect.\ref{sec:Impact of mass dependencies on a mass distribution}), the resulting distribution is a power-law regardless of the initial CMF. In this regime, any turn-over feature is suppressed.
When mass gradients vary with mass dominate and exceed stochastic contributions, a turnover emerges at mass where $\phi'_M = 0$. A power-law tail at high masses is still possible, provided that the ratio $0 < \xi'_M/\phi'_M < 1$ and is constant at high mass (Fig.\ref{fig:convergence_slope}). However, in this case, our scale-free assumption does not hold anymore as two regimes of fragmentation are needed.
When stochastic fragmentation dominates, that is when mass derivatives are small $\phi'_M \approx 0$ and $\xi'_M \approx 0$, the distribution develops log-normal features at its boundaries, with a turn-over at low masses (Sect.\ref{sec:Hierarchical fragmentation applied to a top-heavy CMF}). However, without mass gradients, no power-law is expected at the high-mass end.}

In this scael-free framework, two modes of fragmentation are required to recover the whole shape of the IMF: the peak and the log-normal part of the IMF are determined by random, stochastic processes while the high-mass power-law part of the cIMF is determined by mass-dependent fragmentation. 

We discuss here the possibility for hierarchical fragmentation to shape a Salpeter-like IMF at high masses while setting a high MF for massive primary stars. To set such MF, we found in Sect.~\ref{sec:Multiplicity fraction} that the fragmentation rate needs to be high at high masses and $\phi'_M \geq 0$. Moreover, in the context of hierarchical fragmentation, massive stars with masses $M > 10~$M$_\odot$ can only form if $\xi'_M > 0$ (see Sect.~\ref{sec:Massive star formation}). These two conditions are incompatible with our theoretical results from Sect.~\ref{sec:Convergence to the Salpeter IMF}, stating that both $\xi'_M < 0$ and $\phi'_M < 0$ are required to converge to a Salpeter IMF, unless this IMF is reached coincidentally. From this perspective, a universal IMF cannot emerge from scale-free hierarchical fragmentation alone, accounting only for deterministic mass-dependent processes.

\subsection{Effect of other processes}
\label{Effect of other processes}


\subsubsection{Role of disk fragmentation}
The shape of the IMF may be influenced by fragmentation events occurring on scales $R < R_\text{stop}$, below the core fragmentation stopping scales as disk fragmentation occurs. However, we do not expect disk fragmentation to significantly impact the shape of the fCMF, as a single fragmentation event marginally affects the power-law index. Since few fragmentation steps are expected, a mass dependence of the fragmentation rate ($\phi'_M \neq 0$) should also have little influence, especially if one considers that $\phi(M)$ slowly varies with mass \citep{houghton2024}. At this stage, the primary factor that modifies the shape of the fCMF is through a mass-dependent mass transfer rate ($\xi'_M \neq 0$) for scales $R < R_\text{stop}$, which implies using a variable mass efficiency to map the last fragmenting cores with their stars.

Disk fragmentation also alters the multiplicities of stellar systems by producing close binaries, which could reconcile the MF of high-mass stars with observations. However, if fragmentation continues at scales smaller than $R_{\text{stop}} < 150$~AU, lower-mass cores would form and massive stars ($M > 10$~M$_\odot$) would be even more difficult to form unless $\xi'_M$ is larger as discussed above (see Sect.~\ref{sec:Massive star formation}).

\subsubsection{Impact of the local mass reservoir}
In our model, we assume that the mass transfer rate $\xi(M)$ does not depend on the spatial scale $R$ (scale-free assumption), and that it depend{s} only on the mass $M$ of the parental object. In a larger context in which cores interact dynamically with their environment, accretion processes may occur which can also depend on the gas density from which mass is accreted. From a theoretical perspective, the relationship between the spatial distribution of the fragments (of number density $n \propto r^{-2}$) and the spatial distribution of the gas reservoir (of volumetric density $\rho \propto r^{-2}$) sets the convergence of the power index of the high-mass tail of the IMF towards the value $\Gamma = -1/2$ for gas-dominated potential \citep{bonnell2001}. As long as the mass accretion rate increases with the core mass as $\dot{M} \propto M^x$ with $x > 1$, we expect that $\xi'_M > 0$ since the initial core mass grows faster the more massive it is (otherwise we expect $\xi'_M < 0$). However, because of the fluctuating availability of the local accretion reservoir of the cores that depends on their position in the cloud \citep{Ballesteros-Paredes2015}, it is possible that $\xi'_M < 0$. In addition, when considering temporal aspects, the condition $\xi'_M > 0$ is not necessarily verified if some objects deplete their accretion reservoir before the others. For example, if more massive objects stop accreting at time $t$ while the less massive ones continue to grow \citep{maschberger2014}, we have $\xi'_M < 0$. These couplings between the core mass, the available accretion reservoir, and the duration for which this reservoir remains available could rebalance the high-mass tail of the IMF to -1 \citep{Ballesteros-Paredes2015}. 

\subsubsection{Dynamical interactions}
 {In addition,} core-to-core mutual interactions may also modify the fragmentation properties and induce evolutionary effects. For example, if more massive structures fragment more at time $t_0$ (i.e., $\phi'_M(t_0) > 0$), the resulting gas fragments become more clustered, which may enhance the probability of core coalescence \citep{inutsuka1997}. According to recent numerical simulations, one third of the cores are suspected to experience coalescence, indicating that it is a frequent and important phenomenon \citep{offner2022}. In that case, more massive parents possess more clustered cores that {subsequently} merge more frequently, resulting in comparatively fewer fragments at a later time. Thus, the fragmentation rate effectively satisfies $\phi'_M(t_0 + \Delta t) < 0$. The complex dynamic associated with coalescence gas processing may also induce more gas to be removed from the initial parent, so the mass transfer rate may also effectively satisfy $\xi'_M(t_0 + \Delta t)<0$. {These considerations need to be more deeply quantified to describe the full scenario of fragments formation.}

\subsubsection{Stellar feedback}
Alternatively, stellar feedback from the first-born massive stars can locally prevent further fragmentation \citep{myers2013} and expel gas material from the cores.{In addition, \cite{zhou2025} showed that lower-mass cores in the W43-MM2\&MM3 regions need to have higher mass efficiency in order to match the stellar IMF. This result is compatible with a $\xi'_M<0$ scenario.} This would imply a dependence of the fragmentation properties on the evolutionary stage of the cloud. {In order to have a universal high mass slope,} both conditions $\xi'_M<0$ and $\phi'_M<0$ may arise during the latest stages of core evolution and/or star formation, specifically for the cores hosting the stars of the Salpeter part of the IMF. In that case, if we only account for core hierarchical fragmentation, the Salpeter slope would be intrinsically connected to the physics of massive stars at the latest stages of star formation, while MFs would be a remnant of earlier phases of star formation when no radiative feedback take place, with $\phi'_M > 0$.

\subsubsection{Two phases of fragmentation ?}
Following the above discussion, we propose that star formation occurs in two phases. During the early phases of core evolution, fragmentation occurs with $\phi'_M > 0$ and $\xi'_M > 0$ and sets the multiplicity properties of future stellar systems. Those conditions result in top-heavy CMF with slopes $\Gamma > -1$. Then, as these conditions revert in later stages due to dynamical interactions and stellar feedback, we expect the slope $\Gamma$ to decrease with $\Gamma < -1$.

%
%

\section{Conclusion}
\label{sec:conclusion}

We emphasize that all acronyms and variable names used throughout this work are listed in Appendix~\ref{appendix:acronymTable}, in Tables \ref{tab:acronyms} and \ref{tab:definitions} respectively. We have applied the fragmentation model framework from \cite{thomasson2024} to investigate the scale-free fragmentation of prestellar cores. In particular, we discussed the means to recover both the cIMF and the {observed} multiplicity of stellar systems in scenarios in which only hierarchical fragmentation occurs. To quantify the resulting mass distribution of a stellar population and their clustering as stellar groups, gas cloud is structured on several spatial scales, {equivalent} to fragmentation levels. {We defined}: (i) the fragmentation {spatial} rate $\phi$, which determines the number of fragments produced {at each scale}; (ii) the mass transfer {spatial} rate $\xi$, which describes the efficiency with which parental structure sub-fragments {at each scale}; (iii) the mass partition of the fragments $q > 1$, which regulates the mass distribution between fragments at one level; and (iv) the scale $R_\text{stop}$, below which fragmentation ends.

We distinguished two contributions that influence the shape of the CMF {during} hierarchical fragmentation. First, we investigated the impact of stochastic hierarchical fragmentation with $\xi$ and $\phi$ independent of the mass. We mapped the spatial evolution of a top-heavy CMF extracted from the W43-MM2\&MM3 ridge \citep{pouteau2022} across the scales, starting from $R_0 = 2500$~AU down to varying $R_\text{stop}$. Second, we quantified the slope variations of a CMF, {when the fragmentation rate $\phi(M)$ and the mass transfer rate $\xi(M)$ are dependent on the core mass $M$.}

Using this fragmentation model, we have shown that:

\begin{itemize}
    \item The shift of the CMF towards lower masses is quantified by the fragmentation rate and the mass transfer rate.
    \item {Fragmentation processes that are independent of mass produce IMFs with the same shape, with mass partitions $2<q<4$ and average number of fragments $> 2.5$.}
    \item The resulting IMFs are inconsistent with both the formation of high-mass stars, because of the log-normal cutoff at masses $M > 10~$M$_\odot$; and the multiplicity fractions of stellar systems, as most massive fragments possess five times less companions than the least massive ones. 
    \item A universal Salpeter IMF emerges from hierarchical fragmentation if the fragmentation rate and the mass transfer rate depend on the mass of the initial core, such as $\xi'_M = 0.26 \phi'_M$, with $\xi'_M = \partial \xi / \partial \log M < 0$ and $\phi'_M = \partial \phi / \partial \log M < 0$.
    \item {Multiplicity fractions increase with primary mass if $\phi'_M >0$. Massive stars are likely formed if $\xi'_M >0$.}
\end{itemize}

{The last two statements clashes. Therefore, scale-free hierarchical fragmentation alone cannot simultaneously reproduce both the full shape of the cIMF (at low and high mass), and the observed stellar multiplicity.}

One possibility is that fragmentation is, in fact, not scale-free. In that perspective, we suggest that star formation occurs in two phases. During the initial phase, more massive cores fragment more ($\phi'_M > 0$), with higher mass efficiencies ($\xi'_M > 0$). This phase seeds the multiplicity of the upcoming stellar systems. During the second phase, core dynamics and radiative feedback from the first-born massive stars may prevent further fragmentation (i.e. $\phi'_M < 0$), which may also result in lower mass efficiency (i.e. $\xi'_M < 0$) for more massive cores. In such scenario, the Salpeter slope is determined by mass-dependent processes, while the {low-mass part} of the IMF is shaped by {mass-independent fragmentation  for which} $\phi'_M = 0$ and $\xi'_M = 0$, down to $10^{-2}~M{_\odot}$. 

Alternatively, scale-free fragmentation may reproduce only the cIMF, while the clustering of massive stars would be determined by stellar dynamics after star formation has taken place. In that case, more massive, less clustered stars constitute more stable systems than more clustered, less massive ones. This hypothesis {needs to} be investigated by using the stellar systems formed through scale-free fragmentation as initial conditions for {simulations of cluster dynamical evolution.}

To investigate the origin of the stellar IMF from any CMF, detailed information about individual core properties is necessary to accurately map both mass functions, whether these cores sub-fragment or constitute the last fragmenting structures before the formation of individual stars. We have addressed the theoretical aspect of hierarchical fragmentation, although we lack quantitative multi-scale measurements to assess the mass dependencies of both the fragmentation rate and the mass transfer rate. Nonetheless, we provide a general framework as a basis for such measurements and investigate the influence of fragmentation on the relationship between the shape of the IMF and the underlying stellar clusters. The fragmentation rate and the mass transfer rate can also be used as observable metrics to test our model prediction, and to quantitatively compare core subfragmentation in star-forming regions and numerical simulations.

\begin{acknowledgements} 
BT carried out this project under equal supervision of IJ and EM who both provided comments and suggestions on this manuscript. FM, TY and AG provided helpful discussion and comments on the paper. BT, IJ and FM have received financial support from the French Agence Nationale de la Recherche (ANR) through the project “COSMHIC” (ANR-20-CE31-0009) and the “StarFormMapper” project funded by the European Union’s Horizon 2020 Research and Innovation Action programme (Grant 687528). BT acknowledges the ALMA-IMF consortium (ALMA project \#2017.1.01355.L) for providing data and useful discussion on this work.
\end{acknowledgements}

%
%

\bibliographystyle{aa} 
\bibliography{biblio}

%
%

\appendix

\section{Tables for definitions}
\label{appendix:acronymTable}

In this appendix we show useful tables in which all acronyms (Table \ref{tab:acronyms}) and the relevant model parameters (Table \ref{tab:definitions}) we use in text are listed and defined.

\begin{table}[!h]
    \centering
    \caption{Acronyms used throughout the article.}
    \begin{tabular}{l l l}
    \hline
    \hline
        Acronym & (Section) & Meaning \\
        \hline \\
        MW &(\ref{Section:Introduction}) & Milky Way  \\
        PDF &(\ref{Section:Introduction}) & Probability Density Function \\
        IMF &(\ref{Section:Introduction}) & Initial Mass Function \\
        CMF &(\ref{Section:Introduction}) & Core Mass Function \\
        cIMF &(\ref{Section:Introduction}) & canonical IMF \\
        fCMF &(\ref{sec:Hierarchical fragmentation applied to a top-heavy CMF}) & fragmented Core Mass Function \\
        L3-cIMF &(\ref{sec:Statistical comparison with cIMF}) & Logistic3 IMF \citep{masch2013} \\
        AD &(\ref{sec:Statistical comparison with cIMF}) & Anderson-Darling \\
        MF &(\ref{sec:Stellar systems multiplicity fractions}) & Multiplicity Fraction \\
        fMF &(\ref{sec:Stellar systems multiplicity fractions}) & fragmented Multiplicity Fraction \\
        \hline
    \end{tabular}
    \label{tab:acronyms}
\end{table}

\begin{table*}
    \centering
    \caption{Definitions of the relevant variables used in the model. The `--' character indicates the variable is not referenced in equations since it is introduced as a definition in the text.}
    \begin{tabular}{l l l}
    \hline \hline
        Symbol & Equation & Definition \\[10pt]
        \hline
        $l$ & -- & Level of fragmentation \\
        $R_l$ & -- & Spatial scale at level $l$ \\
        $R_\text{stop}$ & -- & Spatial scale beyond which core fragmentation stops \\
        $i$ & -- & Label that identifies a specific child within one parent when several fragments are produced on the same scale\\
        $M_{l,i}$ & \ref{eq:mass_of_children} & Mass of the i--th child at level $l$ \\
        $\langle M_{l} \rangle$ & \ref{eq:average_mass_scale_free} & Average mass of all fragments at the level $R_l$  \\
        
        $r$ & -- & Scaling ratio between two successive levels \\
        $n_l$ & -- & Number of fragments produced at level $l+1$ by one parent at level $l$ \\
        $p_l(n_l)$ & -- & Probability a parent at level $l$ produces $n_l$ fragments at level $l+1$  \\
        $\bar{n_l}$ & \ref{eq:expecancy_pl} & Expected number of fragments one parent at level $l$ produces at the next level \\

        $\epsilon_l$ & \ref{eq:micro_epsilon_phi} & Mass efficiency with which fragments are formed from their parent at the previous level \\
        $\overline{\epsilon_*}$ & -- & Average mass efficiency with which the last fragmenting cores at $R_\text{stop}$ turn into stars \\
        $q$ & -- & Mass ratio between the more massive child and the less massive one within one siblinghood \\
        $\psi_{l,i}(n_l, q)$ & \ref{eq:omegas_expression} & Fraction of mass received by the i--th children from its parental reservoir $\epsilon_l M_l$ \\

        $\Gamma(R, M)$ & \ref{eq:alpha_pdf} & Power-law index of a mass function in a $\partial N / \partial \log M$ representation at mass $M$ and scale $R$ \\
        $\phi(M)$ & \ref{eq:micron_phi} & Spatial fragmentation rate  \\
        $\phi'_M$ & -- & Derivative of the fragmentation rate with respect to the mass \\

        $\xi(M)$ & \ref{eq:micro_epsilon_phi} & Spatial mass transfer rate \\
        $\xi'_M$ & -- & Derivative of the mass transfer rate with respect to the mass \\
        $\mathcal{E}(R)$ & -- & Fragment formation mass efficiency between the cloud and the cores at scale $R$ \\
        \hline
    \end{tabular}
    \label{tab:definitions}
\end{table*}

\section{Variation of number of fragments}
\label{appendix_demo_slopevar}


The purpose of this appendix is to derive Eq.~\ref{eq:final_slope_var} we introduce in Sect.~\ref{sec:Impact of mass dependencies on a mass distribution} that characterises the variation of the local power index $\Gamma(R, M)$ after a serie of fragmentation events throughout the spatial scales $R$, at mass $M$. Consider the number of fragments $N(R, M)$ of mass $M$ and size $R$. In the logarithmic space, it is always possible to locally characterise such a function from the slope $\Gamma(R, M)$ which represents the tangent line of the function $\log N(R, M)$ with respect to its logarithmic in of mass $\log M$ by

\begin{equation}
    \Gamma(R, M) = \dfrac{\partial \log N(R, M)}{\partial \log M}\Big|_R
    \label{eq:def_gamma}
.\end{equation}

\noindent Assuming the mass variations of $\Gamma(R, M)$ are small in an interval $d \log M$, we can locally approximate the function $N(R, M)$ by $N(R, M) \propto M^{\Gamma}$. On the other hand, the mass distribution at one spatial scale $R$ is defined as the number of fragments counted within a finite mass interval and we can show that

\begin{equation}
    \dfrac{\partial N(R, M)}{\partial \log M} \propto M^{\Gamma}
.\end{equation}

\noindent Therefore, the power index $\Gamma(R, M)$ describes the number of fragments within a small mass interval of $d \log M$. As long as we consider local variations in mass, this description remains valid. For example, the Salpeter power-law index $\Gamma = 1.35$ quantifies the local variations of the IMF within the $M > 1~$M$_\odot$ mass range.

To assess the variations of the local index $\Gamma(R, M)$ at mass $M$ along fragmentation events that cascade through spatial scales $R$, we take the following derivative with Eq.~\ref{eq:def_gamma}

\begin{equation}
    \dfrac{\partial \Gamma(R, M)}{\partial \log R} = \dfrac{\partial}{\partial \log R} \left(  \dfrac{\partial \log N}{\partial \log M} \right)  = \dfrac{\partial}{\partial \log M} \left( \dfrac{\partial \log N}{\partial \log R} \right)
    \label{app:eq:dgammadR}
,\end{equation}

\noindent where we use the symmetry of second derivatives. We then need to compute the variation of the number of fragments of mass $M$ between two successive scales. 

\begin{figure}
    \centering
    \includegraphics[width=9cm]{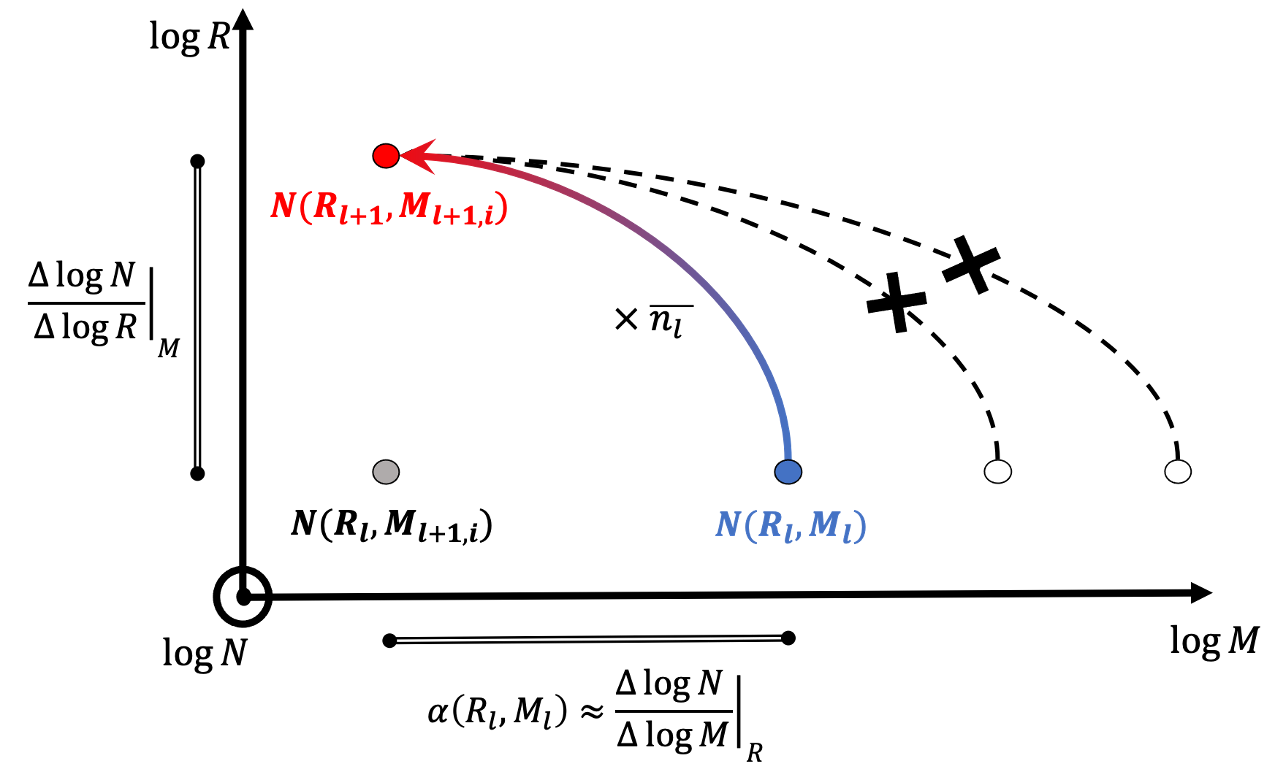}
    \caption{Schematic view of the number of children of mass $M_{l+1}$ created at the scale $R_{l+1}$ (red dot) from one parental object of mass $M_l$ at scale $R_l$ (blue dot). In this work we assume that children result from the outcome of a single parental mass (solid coloured line) and neglect other possible contributions (crossed dotted lines). Since only the blue dot injects children in the red dot, the number of fragments inside the red dot corresponds to the number of parents inside the blue dot multiplied by the average number of fragments $\overline{n_l}$ created by these parents.}
    \label{app:fig:schematic_deltaNdeltaR}
\end{figure}

We assume that only the parental objects of mass $M_p$ can form children of mass $M$. Hence, the number of children produced at scale $R_{l+1}$ corresponds to the number of parents at scale $R_{l}$ (with mass $M_p$) multiplied by their average number of children $\overline{n_l}$ created (see Fig.~\ref{app:fig:schematic_deltaNdeltaR}):

\begin{equation}
    N(R_{l+1}, M) = \overline{n_l} N(R_{l}, M_p)
    \label{app:eq:averageNchild}
.\end{equation}

We also recall Eq.~\ref{eq:micron_phi} that quantifies the average number of fragments produced $\bar{n_l}$ between two scales by one parent with the fragmentation rate $\phi$:

\begin{equation}
    \bar{n_l} = \left( \frac{R_{l+1}}{R_{l}} \right )^{-\phi}
    \label{app:eq:recallfrag}
.\end{equation}

\noindent This equation can be extended by considering the individual fragmentation rate $\phi(M)$ that depends on the mass of the parental object.

Next, we approximate the derivative of $\partial \log N / \partial \log R$ as

\begin{equation}
    \dfrac{\partial \log N}{\partial \log R}\Big|_M = \dfrac{\log N(R_l, M) - \log N(R_{l+1}, M)}{\log R_l - \log R_{l+1}}
.\end{equation}

\noindent Using Eq.~\ref{app:eq:averageNchild} we obtain

\begin{equation}
    \dfrac{\partial \log N}{\partial \log R} = \dfrac{\log N(R_l, M) - \log N(R_{l}, M_p) - \log \bar{n_l}}{\log R_l - \log R_{l+1}}
,\end{equation}
    
\noindent which can be written as

\begin{equation}
    \begin{split}
    \dfrac{\partial \log N}{\partial \log R} & = \dfrac{\log N(R_l, M) - \log N(R_{l}, M_p)}{\log M - \log M_p} \times \dfrac{\log M - \log M_p}{\log R_{l} - \log R_{l+1}} \\ &  - \dfrac{\log \bar{n_l}}{\log R_{l} - \log R_{l+1}}
    \end{split}
.\end{equation}
    
\noindent We recognise the first term as being the local slope $\Gamma(R, M)$ and the third term can be simplified using Eq.~\ref{app:eq:recallfrag} as

\begin{equation}
     \dfrac{\partial \log N}{\partial \log R} = \Gamma \times \dfrac{\log M - \log M_p}{\log R_{l} - \log R_{l+1}} - \phi
.\end{equation}
   
\noindent Since the parental mass $M_p$ lies at the spatial scale $R_{l+1}$ while the children mass $M$ lies at the scale $R_l$, we can write

\begin{equation}
    \dfrac{\partial \log N}{\partial \log R} = - \Gamma \delta_M - \phi
    \label{app:eq:dNdR}
,\end{equation}


\noindent where $\delta_M = \dfrac{\Delta \log M}{\Delta \log R}$ is the mass variation between a parent its child of mass $M$ between two successive levels of fragmentation. The component $\phi$ reflects the fact that multiple fragments of mass $M$ come from the same parental mass. Then, we can substitute Eq.~\ref{app:eq:dNdR} into Eq.~\ref{app:eq:dgammadR} so

\begin{equation}
    \dfrac{\partial \Gamma(R, M)}{\partial \log R} = - \dfrac{\partial}{\partial \log M} \big( \Gamma \delta_M \big) - \dfrac{\partial \phi}{\partial \log M}
    \label{app:eq:general_alphavar}
.\end{equation}

\noindent Within the framework of our model, the mass variation $\delta_M$ between a parent and a child is given using Eq.~\ref{eq:mass_of_children} by

\begin{equation}
    \delta_M = \dfrac{\log \epsilon_l \psi_{l,i}(n_l)}{\log R_{l+1} - \log R_{l}} 
    \label{eq:pace}
,\end{equation}

\noindent so that the pace $\delta_M$ quantifies by how much mass one fragment is shifted from its original parental mass bin through the spatial scales.

Assuming that all the children are formed with the same mass efficiency, we can consider an average slope variation and $\delta_M \approx \phi - \xi$ (see Eq.~\ref{eq:average_mass_scale_free}) which reads 

\begin{equation}
    \dfrac{\partial \Gamma(R, M)}{\partial \log R} + \big(\phi - \xi) \dfrac{\partial \Gamma(R, M)}{\partial \log M} = \Gamma \dfrac{\partial \xi(M)}{\partial \log M} - (1 + \Gamma) \dfrac{\partial \phi(M)}{\partial \log M}
    \label{app:eq:mean_alphavar}
.\end{equation}

\noindent This equation is analogous to an advection equation in which the local slope $\Gamma(R, M)$ is transported throughout the mass domain $M$ with the successive fragmentation events across the spatial scales $R$ at a pace $\phi - \xi$. Two additional terms describe the influence of mass dependencies on the evolution of the local slope $\Gamma(R, M)$.

\section{Procedural generation of a fCMF}
\label{app:Procedural generation of a fCMF}

We derive the Eq.~\ref{eq:zeta_l_expression} used to procedurally compute the fCMF at a specific spatial scale in Appendix~\ref{app:sec:Derivation of Eq.} , and we compare the resutling fCMF with the log-normal CMF of \cite{larson1973_fragmentation} and a Monte-Carlo sampling in Appendix~\ref{app:sec:Reliability of our procedure} to check the validity of our method.

\subsection{Derivation of Eq.~\ref{eq:zeta_l_expression}}
\label{app:sec:Derivation of Eq.}

In order to hierarchically fragment a mass distribution and derive the resulting fCMF, we introduce a semi-analytic procedural method. The following method remains valid assuming (i) $\epsilon_{l}$ is not a random variable and (ii) both the fragmentation rate $\phi$ and mass transfer rate $\xi$ do not depends on the mass of the parental object. The mass distribution of a population of fragments located inside any level $l$ is described by a PDF $\zeta_l(M)$ normalised as

\begin{equation}
    \displaystyle \int_{0}^{+\infty} \zeta_l(M) dM = 1
    \label{eq:Chap5:normzeta}
.\end{equation}

The mass function $\zeta_{l+1}(M)$ associated to the population of the next level can be derived from $\zeta_{l}(M)$ by considering every possible fragmentation outcome of the parents constituting $\zeta_{l}(M)$. These parents can be grouped into different sub-populations, depending on the amount of children $n_l$ they produce and the fraction of mass $\epsilon_l \psi_{l,i}(n_l)$ each individual children receive from their associated parent. For example, all the $i-th$ children originating from a $n_l = 2$ outcome who have received the fraction $\epsilon_l \psi_{l,i}(n_l = 2)$ from their parent, constitute one sub-population of $\zeta_{l+1}(M)$. At level $l+1$, a sub-population is then characterised by the collection of fragments originating from the same pair ($n_l$ ; $\epsilon_l \psi_{l,i}$). 

On one hand, the amount of children within a siblinghood of $n_l$ fragments is $N_\text{tot}(R_l) \times n_l \times p_l$, where $N_\text{tot}$ is the total number of parents at scale $R_l$. Since there is exactly $n_l$ children in the siblinghood, the number of children that are produced with the efficiency $\epsilon_l \psi_{l,i}$ in such siblinghood is $N_\text{tot}(R_l) \times n_l \times p_l / n_l$. This is the amount of children constituting the sub-population associated to the pair ($n_l$; $\epsilon_l \psi_{l,i}$). On the other hand, the total amount of children produced considering every possible outcome is $N_\text{tot}(R_l) \times \bar{n_l}$, where $\bar{n_l}$ is given in Eq.~\ref{eq:expecancy_pl} as the expected value of the probability distribution $p_l$. The probability $\nu_{l}$ of randomly drawing a fragment from the sub-population characterised by the pair ($n_l$; $\epsilon_l \psi_{l,i}$) is the ratio between the amount of children constituting this sub-population with the total amount of children produced considering every outcome, which is

\begin{equation}
    \nu_{l}(n_l) = \dfrac{p_l(n_l)}{\bar{n_l}}
    \label{eq:probability_subpopulation}
.\end{equation}

\noindent We check the probability $\nu_{l}(n_l)$ is normalised by summing over all of the possible sub-populations, associated with the possible outcome pairs ($n_l$ ; $\epsilon_l \psi_{l,i}$):

\begin{equation}
    \displaystyle \sum_{n_l} \sum_{i=1}^{n_l} \nu_{l}(n_l) = \sum_{n_l} \dfrac{p_l(n_l)}{\bar{n_l}} \sum_{i=1}^{n_l} 1 = \frac{1}{\bar{n_l}} \sum_{n_l} p_l(n_l) n_l = 1
.\end{equation}

The mass of the children constituting any sub-population is related to the mass of their parent by $M_{l+1, i}(n_l) = \epsilon_l \psi_{l,i}(n_l) M_{l}$. Therefore, to generate the corresponding sub-population at scale $l+1$ the mass of the parents at scale $l$ is reduced by a factor $\epsilon_l \psi_{l,i}(n_l)$. We can write the expression of the PDF $\zeta_{l+1}(M)$ by summing the mass contributions of each of the sub-populations weighted by their probability of occurrence so

\begin{equation}
    \zeta_{l+1}(M) = A \displaystyle \sum_{n_l} \sum_{i=1}^{n_l} \nu_{l}(n_l) \zeta_{l}\left(\dfrac{M}{\epsilon_l \psi_{l,i}(n_l)}\right)
,\end{equation}

\noindent where we can show that $A = \dfrac{\bar{n_l}}{\epsilon_l}$ is the normalisation coefficient and we finally obtain

\begin{equation}
    \zeta_{l+1}(M) = \displaystyle \sum_{n_l} \sum_{i=1}^{n_l} \dfrac{p_l(n_l)}{\epsilon_l} \zeta_{l}\left(\dfrac{M}{\epsilon_l \psi_{l,i}(n_l)}\right)
    \label{app:eq:zeta_l_expression}
.\end{equation}

\subsection{Reliability of our procedure}
\label{app:sec:Reliability of our procedure}

The scale-free model we introduce is a generalisation of the stochastic model developed by \cite{larson1973_fragmentation}, later L73. Both of these models formalise the fragmentation process along several discrete fragmentation levels in which each parent randomly fragments at the next level. In L73's model, an object of mass $M_0$ at the initial level $l = 0$ has a probability $p$ of fragmenting into two children of equal mass $M_0/2$ towards the level $l = 1$, and a probability $1 - p$ of collapsing into a single child of mass $M_0$. After $l$ levels of fragmentation, the children can only acquire discrete mass values $M_n = 2^{-n} M_0$, where $n \leq l$ is the number of fragmentation events that have occured after $l$ levels. With their model, L73 shows that the fraction $f(n, l)$ of the cloud mass falling into fragments of mass $M_n = 2^{-n} M_0$ after $l$ fragmentation levels is given by the binomial distribution

\begin{equation}
    f(n, l) = p^n (1-p)^{n-l} \frac{n!}{l ! (n - l)!}
.\end{equation}

If $N(M_n)$ is the number of fragments inheriting a mass $M_n$, the fraction $f(n, l)$ can be written by definition as

\begin{equation}
    f(n, l) = \frac{N(M_n) M_n}{M_0} 
.\end{equation}

Hence, the number of fragments of mass $M_n$ after $l$ levels of fragmentation is

\begin{equation}
    N(M_n) = 2^n f(n, l)
.\end{equation}

With the previous equation, we can obtain the discrete distribution of the fragments mass. In their model, L73 uses the central limit theorem to derive the approximated continuous probability density of this mass function. To assess the correctness of our procedural approach, we compare the discrete mass distribution $s(M_n)$ of L73 with our fCMF derived from the semi-analytical solution expressed by Eq.~\ref{eq:zeta_l_expression}, under the same fragmentation conditions as in L73. 

\begin{figure}
    \centering
    \includegraphics[width=9cm]{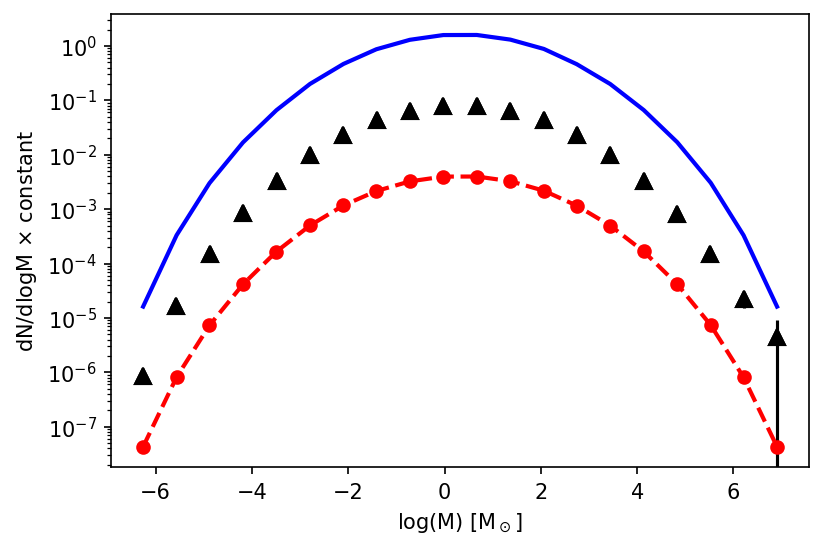}
    \caption{Mass functions derived from our scale-free fragmentation model (blue), \cite{larson1973_fragmentation} (L73) model (red) and a Monte-Carlo procedure (black). The mass functions are computed under the same condition as L73 with a pseudo-Dirac as a starting mass distribution centered at $10^3{\rm M}_\odot$, 19 levels of fragmentation, mass transfer rate $\xi = 0$ and fragments mass ratio $q = 1$. At each level of fragmentation the objects subdivide into one or two fragments with respective probabilities 0.5 and 0.5. The different mass functions are multiplied by different constants to separate them for better visibility.}
    \label{fig:comparison_larson}
\end{figure}

In the framework of our scale-free model, L73's fragmentation corresponds to equipartition binary fragmentation with $q = 1$, and $\epsilon_l = 100\%$ corresponding to $\xi = 0$. Since the spatial scales of the fragments are not defined in L73, an equivalent fragmentation rate $\phi$ cannot be estimated. Hence, we set $\phi = 1$ and adapt a scaling ratio $r = 1.5$ so that a parent generates either $n = 1$ or $n = 2$ children each with a probability $p = 0.5$ using Eq.~\ref{eq:probability_phi}. We start from an initial mass distribution described by a normal distribution whose standard deviation tends to 0 in order to simulate a Dirac distribution, centred in $M_0 = 10^3 {\rm M}_\odot$. We also compare these distributions with the mass function resulting from a Monte-Carlo procedure in which we randomly fragment individually $N = 10,000$ objects of mass $m_0 = 10^3 {\rm M}_\odot$. This Monte-Carlo solution is supposed to be an approximation of the true mass distribution of the fragments at the level we consider, here $l = 19$ as L73. Both the L73 and our semi-analytical solution match the Monte-Carlo distribution (see Fig.~\ref{fig:comparison_larson}). We can be confident of using our procedural approach. 

\section{Mass dependent model validity}
\label{app:Mass dependent model validity}

In this appendix we check the direction of variation of the local slope $\Gamma(R, M)$ as predicted by Eqs.~\ref{eq:alpha_variation_xi} and \ref{eq:alpha_variation_phi}, depending on the signs of $\partial \phi / \partial \log M$, $\partial \xi / \partial \log M$ and $\Gamma$. In particular, these equations predict respectively that $\Gamma$ increases during fragmentation if:

\begin{itemize}
    \item $\partial \xi / \partial \log M$ have the opposite sign of $\Gamma$;
    \item $\partial \phi / \partial \log M$ have the same sign as $1 + \Gamma$.
\end{itemize}

\noindent Inversely, $\Gamma$ decreases during fragmentation if:

\begin{itemize}
    \item $\partial \xi / \partial \log M$ have the same sign as $\Gamma$;
    \item $\partial \phi / \partial \log M$ have the opposite sign of $1 + \Gamma$.
\end{itemize}

\noindent We also have the following conditions:

\begin{itemize}
    \item if $\partial \xi / \partial \log M > 0$, $\Gamma(R)$ converges to 0;
    \item if $\partial \phi / \partial \log M < 0$, $\Gamma(R)$ converges to -1.
\end{itemize}

To check these statements, we compare the expected slope predicted by Eqs.~\ref{eq:alpha_variation_xi} and \ref{eq:alpha_variation_phi} we recall in Appendix~\ref{app:sec:Theoretical variation of alpha} with the slope of a simulated fragmented CMF (more details in Appendix~\ref{app:sec:Fragmentation simulation}). We measure the slope of the fragmented distribution by fitting it with a power-law function and we compare this fit with the theoretical prediction given in the following. 

\subsection{Theoretical variation of $\Gamma(R)$}
\label{app:sec:Theoretical variation of alpha}

We compute the theoretical variation of $\Gamma(R)$ associated with the mass derivative of $\xi(M)$ or $\phi(M)$ by solving the following differential equations:

\begin{equation}
    \begin{cases}
      \dfrac{\partial \Gamma}{\partial \log R} = \Gamma \xi'_M \\
      \text{variation of $\Gamma$ associated to $\xi(M)$}\\
      \\
      \dfrac{\partial \Gamma}{\partial \log R} = - (1 + \Gamma) \phi'_M 
      \\
      \text{variation of $\Gamma$ associated to $\phi(M)$}
    \end{cases}\,
    \label{appendix:eq:equa_diff_alpha}
,\end{equation}

\noindent where both $\xi'_M$ and $\phi'_M$ are assumed constant for simplicity and satisfy

\begin{equation}
    \begin{cases}
      \dfrac{\partial \xi}{\partial \log M} = \xi'_M\\
      \\
      \dfrac{\partial \phi}{\partial \log M}  = \phi'_M
    \end{cases}\,
    \label{appendix:phi_xi_eq}
.\end{equation}

Solving Eqs.~\ref{appendix:eq:equa_diff_alpha}, we obtain

\begin{equation}
    \begin{cases}
      \Gamma(R) = \Gamma_0 \left( \dfrac{R}{R_0} \right)^{\xi'_M} \\
      \text{variation of $\Gamma$ associated to $\xi(M)$}\\
      \\
      \Gamma(R) = (\Gamma_0 + 1) \left( \dfrac{R}{R_0} \right)^{-\phi'_M} - 1
      \\
      \text{variation of $\Gamma$ associated to $\phi(M)$}
    \end{cases}\,
    \label{app:eq:alpha_full_solution}
,\end{equation}

\noindent where $\Gamma_0$ is the initial slope at scale $R_0$. The two previous equations represent the theoretical variations of $\Gamma(R)$ associated to either the derivative of the mass transfer rate $\partial \xi / \partial \log M$ or the derivative of the fragmentation rate $\partial \phi / \partial \log M$. 

\subsection{Fragmentation simulation}
\label{app:sec:Fragmentation simulation}

We employ a Monte-Carlo method to sample $10^6$ cores within a power-law distribution of initial slope $dN/d\log M = \Gamma_0$ at an initial spatial scale $R_0$. In these Monte-Carlo simulation we consider a fragmentation process with constant $\xi'_M$ and $\phi'_M$. To attribute the mass of the fragments produced and their number, we compute the mass $M$ of the fragments and the expected number of fragments $N$ produced at any scale $R$ given in \cite{thomasson2024} by

\begin{equation}
      \dfrac{d \log M}{d \log R} = \phi(M) - \xi(M)
      \label{appendix:eq:M}
,\end{equation}

\begin{equation}
    \dfrac{d \log N}{d \log R} = - \phi(M)
    \label{appendix:eq:N}
.\end{equation}

Then, we compute $\xi(M)$ and $\phi(M)$ by solving Eqs.~\ref{appendix:phi_xi_eq}:

\begin{equation}
    \begin{cases}
      \xi(M) = \xi'_M \log \left( \dfrac{M}{\Tilde{M_\xi}} \right)\\
      \\
      \phi(M)  = \phi'_M \log \left( \dfrac{M}{\Tilde{M_\phi}}  \right)
    \end{cases}\,
    \label{appendix:solution_xi_phi}
,\end{equation}

\noindent where $\Tilde{M_\xi}$ and $\Tilde{M_\phi}$ represent the masses for which $\xi(M) = 0$ and $\phi(M) = 0$ respectively. 

The previous Eqs.~\ref{appendix:eq:M}, \ref{appendix:eq:N} and \ref{appendix:solution_xi_phi} are used to investigate the influence of mass dependent mass transfer rate $\xi(M)$ and fragmentation rate $\phi(M)$ in the Appendixes \ref{app:sec:Variation of a associated to xi} and \ref{app:sec:Variation of a associated to phi} respectively. In these appendixes, we first check the consistency of our theoretical prediction with our Monte-Carlo sampling regarding the dependance of Eq.~\ref{app:eq:alpha_full_solution} on the initial slope $\Gamma_0$ that can vary between $-1.5$ and $1.5$ and evaluate the mass distributions at a scale $R_0/2$. Then, we check the evolution of $\Gamma$ with the spatial scale $R$ according to Eq.~\ref{app:eq:alpha_full_solution} for initial conditions $\Gamma_0 = -1, 0$ and $1$. 

\subsubsection{Variation of $\Gamma$ associated to $\xi(M)$}
\label{app:sec:Variation of a associated to xi}

To evaluate the variation of $\Gamma$ associated to the variation of $\xi(M)$ with the mass $M$, we assume no fragmentation occurs, that is $\phi(M) = 0$. Injecting Eqs.~\ref{appendix:solution_xi_phi} into Eq.~\ref{appendix:eq:M}, we obtain the following differential equation:

\begin{equation}
    \dfrac{d \log M}{d \log R} = - \xi'_M \log \left( \dfrac{M}{\Tilde{M_\xi}} \right)
.\end{equation}

\noindent Solving this equation we find

\begin{equation}
    \log \left( \dfrac{M}{\Tilde{M_\xi}}\right) = \log \left( \dfrac{M_0}{\Tilde{M_\xi}} \right) \left( \dfrac{R}{R_0} \right)^{-\xi'_M}
    \label{appendix:mass_for_xi}
,\end{equation}

\noindent where $M_0$ is the initial mass of an object at scale $R_0$. For each core, we compute their mass using Eq.~\ref{appendix:mass_for_xi} for any spatial scale $R$. We consider an ad-hoc mass transfer rate law where $\Tilde{M}_\xi$ corresponds to the minimum mass of the sample and $\xi'_M = -1, 0$ or $1$ to respectively test the three cases $\partial \xi / \partial \log M < 0$, $= 0$ or $> 0$. The slopes measured in our Monte-Carlo distributions are compatible with our theoretical model (see Fig.~\ref{app:fig:alpha_alpha0}a) within the range of parameters we evaluated.

The following statements are validated (Fig.~\ref{app:fig:alpha_alpha0}a):

\begin{itemize}
    \item if $\partial \xi / \partial \log M$ have the opposite sign of $\Gamma$, then $\Gamma$ increases and inversely;
    \item if $\partial \xi / \partial \log M > 0$, $\Gamma(R)$ converges to 0.
\end{itemize}

For smaller spatial scale $R \ll R_0$ the Monte-Carlo simulation agrees with the theoretical prediction as $\Gamma(R)$ converges to 0 if $\partial \xi / \partial \log M > 0$ (Fig.~\ref{app:fig:alpha_alpha0}b)

\subsubsection{Variation of $\Gamma$ associated to $\phi(M)$}
\label{app:sec:Variation of a associated to phi}

To evaluate the variation of $\Gamma$ associated to the variation of $\phi(M)$ with the mass $M$, we assume mass conservation for every object, that is $\xi(M) = 0$. Injecting Eqs.~\ref{appendix:solution_xi_phi} into Eq.~\ref{appendix:eq:M}, we obtain the following differential equation for the mass:

\begin{equation}
    \dfrac{d \log M}{d \log R} = \phi'_M \log \left( \dfrac{M}{\Tilde{M_\phi}} \right)
,\end{equation}

\noindent that is solved by

\begin{equation}
    \log \left( \dfrac{M}{\Tilde{M_\phi}} \right) = \log \left( \dfrac{M_0}{\Tilde{M_\phi}} \right) \left( \dfrac{R}{R_0} \right)^{\phi'_M}
    \label{appendix:eq:mass_phi}
.\end{equation}

\noindent On the other hand, we also have the following equation for the expected number of fragments:

\begin{equation}
    \dfrac{d \log N}{d \log R} = - \phi'_M \log \left( \dfrac{M}{\Tilde{M_\phi}} \right)
    \label{appendix:eq:number_phi}
.\end{equation}

Substituing Eq.~\ref{appendix:eq:mass_phi} into Eq.~\ref{appendix:eq:number_phi}, we compute the solution

\begin{equation}
    \log \left( \dfrac{N}{N_0} \right) = \log \left( \dfrac{M_0}{\Tilde{M_\phi}} \right) \left[ 1 - \left( \dfrac{R}{R_0} \right)^{\phi'_M} \right]
    \label{appendix:eq:N_frag}
,\end{equation}

\noindent where $N_0 = 1$ considering that in our Monte-Carlo process, each core at the initial scale $R_0$ is taken individually to fragment on its own depending on its mass. Nonetheless, the previous equation gives the expected number of fragments to be produced. To limit the effect of stochasticity (i.e. a natural steepening of the slope, see Sect.~\ref{sec:Retrieving the L3-cIMF from the fCMF}), we arbitrarily choose to consider a binary fragment distribution in which $\left \lfloor{N}\right \rfloor$ and $\left \lfloor{N}\right \rfloor + 1$ fragments are produced with a probability $1 - \{N\}$ and $\{N\}$ respectively, where $\left \lfloor{N}\right \rfloor$ denotes the integer part of $N$ and $\{N\}$ its fractional part. We also consider a single level of fragmentation so the global shape of the initial distribution is not sufficiently altered by stochasticity.
 
 \begin{figure}
     \centering
     \includegraphics[width=9cm]{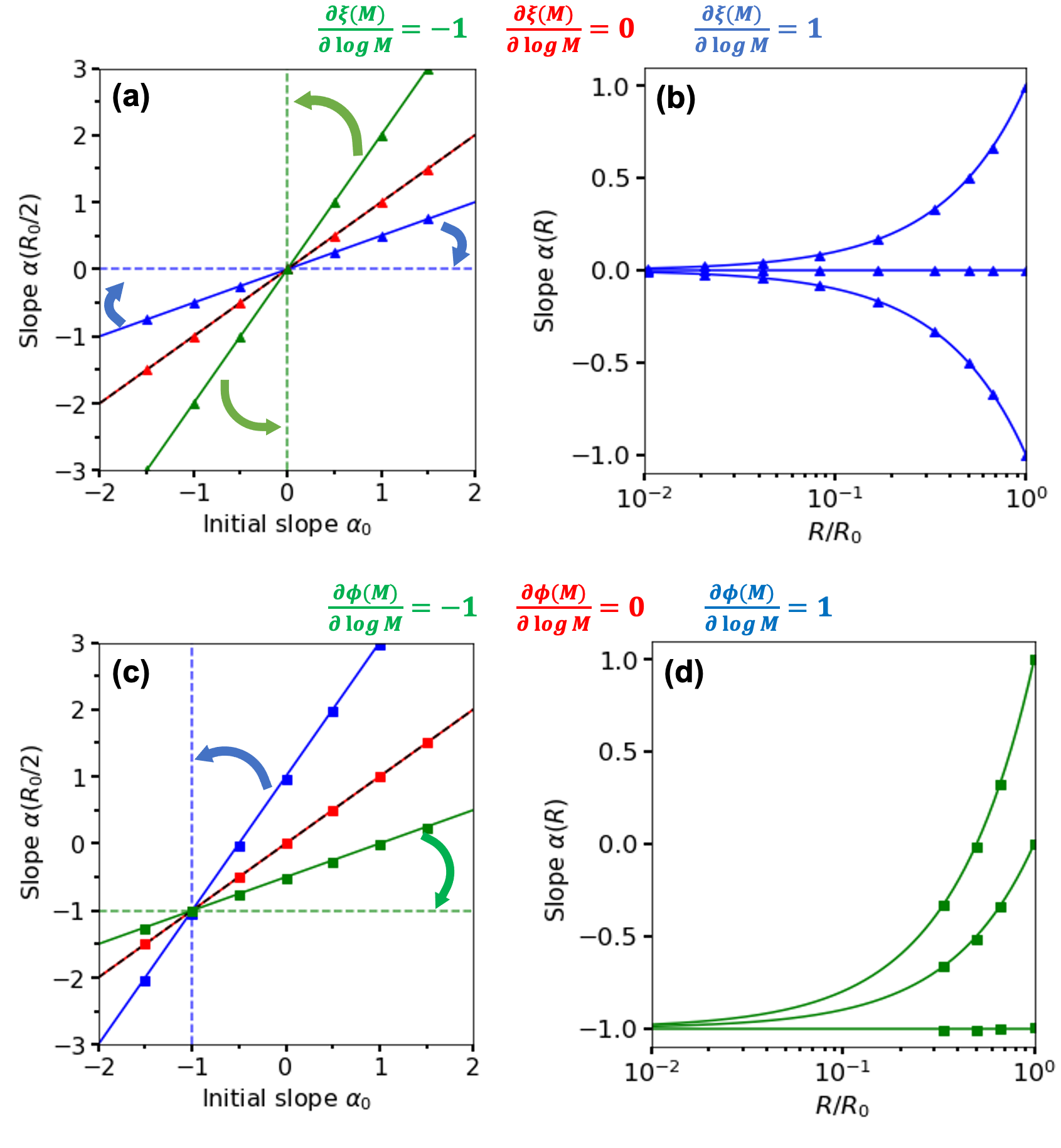}
     \caption{Evolution of the local power-law index $\Gamma$ of a mass distribution. The influence of the mass transfer rate $\xi$ and the fragmentation rate $\phi$ dependencies with the mass $M$ of the parental fragment is shown in (a), (b) and (c), (d) respectively. The solid lines, triangles and squares represent respectively the theoretical solutions from Eq.~\ref{app:eq:alpha_full_solution}, the Monte-Carlo solutions obtained from a sample of $10^6$ objects distributed in a power-law of index $\Gamma_0$ fragmenting according to Eq.~\ref{appendix:mass_for_xi} and Eq.~\ref{appendix:eq:N_frag}. (a)-(c): Local power-law index $\Gamma$ as a function of the initial slope $\Gamma_0$ at scale $R_0/2$. Green and blue dashed lines indicate the asymptotic limits of $\Gamma(R)$. (b)-(d): Local power-law index $\Gamma$ as a function of the spatial scale $R$ for different initial slopes. The Monte-Carlo solutions do not go below $R/R_0 \approx 0.3$ because the number of fragment constituting the fCMF becomes too large to compute.}
     \label{app:fig:alpha_alpha0}
 \end{figure}

For each core of our initial sample, we compute the number of fragments it produces using Eq.~\ref{appendix:eq:N_frag} for any spatial scale $R$. Then, we split the mass of this parental core equivalently between the children and fit the slope of the resulting mass distribution to compare with the predicted slope at this scale using Eq.~\ref{app:eq:alpha_full_solution}. We consider an ad-hoc fragmentation rate function where $\Tilde{M}_\phi$ corresponds to the minimum mass of the sample when $\phi'_M = 1$ or $0$ and $\Tilde{M}_\phi$ corresponds to the maximum mass of the sample when $\phi'_M = -1$ in order to test the three cases $\partial \phi / \partial \log M > 0$, $= 0$ or $< 0$ respectively. The Monte-Carlo fragmented distributions are compatible with our theoretical model (see Fig.~\ref{app:fig:alpha_alpha0}-(b)) within the range of parameters we evaluated. We are confident that the slope variations predicted by our theoretical model represent the actual variations of slope under similar circumstances. 

The following statements are validated (Fig.~\ref{app:fig:alpha_alpha0}c):

\begin{itemize}
    \item if $\partial \phi / \partial \log M$ have the same sign as $\Gamma$, then $\Gamma$ increases and inversely;
    \item if $\partial \phi / \partial \log M < 0$, $\Gamma(R)$ converges to -1.
\end{itemize}

For smaller spatial scale $R \ll R_0$ the Monte-Carlo simulation agrees with the theoretical prediction as $\Gamma(R)$ converges to -1 if $\partial \phi / \partial \log M < 0$ (Fig.~\ref{app:fig:alpha_alpha0}d). 

\end{document}